# Ergodic Capacity of Cognitive Radio under Imperfect Channel State Information

Zouheir Rezki *Member, IEEE,* and Mohamed-Slim Alouini, *Fellow, IEEE,*

*Abstract*—A spectrum-sharing communication system where the secondary user is aware of the instantaneous channel state information (CSI) of the secondary link, but knows only the statistics and an estimated version of the secondary transmitter-primary receiver (ST-PR) link, is investigated. The optimum power profile and the ergodic capacity of the secondary link are derived for general fading channels (with continuous probability density function) under average and peak transmit-power constraints and with respect to two different interference constraints: an interference outage constraint and a signal-to-interference outage constraint. When applied to Rayleigh fading channels, our results show, for instance, that the interference constraint is harmful at high-power regime in the sense that the capacity does not increase with the power, whereas at low-power regime, it has a marginal impact and no-interference performance corresponding to the ergodic capacity under average or peak transmit power constraint in absence of the primary user, may be achieved.

## I. Introduction

Cognitive radio (CR) techniques have been proposed to efficiently use the spectrum through an adaptive, dynamic, and intelligent process [1]. Spectrum utilization can be improved by permitting a secondary user (who is not being serviced) to access a spectrum hole unoccupied by the primary user, or to share the spectrum with the primary user under certain interference constraints [2]. CR refers to different approaches to this problem that seek to overlay, underlay, or interweave the secondary user's signals with those of the primary users [3]. In the underlay settings, cognitive users can communicate as long as the interference caused to non cognitive users is below a certain threshold. Overlay systems, on the contrary, adopts a less conservative policy by permitting cognitive and non cognitive users to communicate simultaneously exploiting side information and using sophisticated coding techniques [4]. Perhaps the most conservative of the three, is the interweave system that permits to cognitive users to communicate provided that the actual spectrum is unoccupied by non cognitive users. More details on these three systems can be found, for instance, in [3], [4]. From an information-theoretical point-of-view, establishing performance limits of these systems relies strongly on the available side information that a cognitive user has about the network nodes: channel state information (CSI), coding techniques, codebooks, etc., e.g., [5]–[8].

In this paper, we focus on a spectrum-sharing CR model under general fading channels, with continuous probability density functions (p.d.f.), where the primary and the secondary users share the same spectrum under certain interference constraints. More specifically, we aim at analyzing the optimal power allocation and the ergodic capacity of the secondary link under limited channel knowledge at the secondary transmitter [9], [10].

Previous works have studied the impact of fading on the secondary link capacity under average or peak transmit-power, but assuming that the secondary transmitter is aware of the instantaneous CSI of the secondary transmitter-primary receiver (ST-PR) link, e.g., [11]–[14]. Although the later assumption generally guarantees an instantaneous limitation of the interference at the primary receiver, it is quite strong to obtain such valuable CSI in absence of an established cooperation protocol between the primary and the secondary links. Recall that protecting the primary user against interference may not be accurate if the CSI needed to estimate interference levels is coarsely precise, as shown in [15]. A step forward to address the problem in a more practical setting considering imperfect CSI has been realized in [16], [17], where the capacity or a lower bound on it has been derived under average received power or average interference outage constraint, respectively; but neither an average nor a peak transmit-power has been considered. The effect of ST-PR channel estimation at the SU on the ergodic capacity has also been analyzed under peak transmit power and peak interference constraint at the primary receiver, in [18]. However, unless some assumptions on the interference (strong or weak) caused by the primary user at the secondary receiver are adopted in the more general interference channel model therein, the results obtained seem to be an achievable rate using Gaussian codebook, as the capacity of the interference channel is still generally not known. Along similar lines, [19] considers the effect of statistical CSI rather than instantaneous channel estimation errors. Likewise, a rate-maximization problem of a secondary link where the cognitive users are equipped with multiple antennas and under an average transmit power along with an average interference constraints at the primary receiver has been considered in [20]. The secondary transmitter has been assumed to know the mean or the covariance of the ST-PR CSI through feedback, and in both cases, algorithms have been proposed to find the instantaneous optimum rate. In [21], a sum-rate maximization of the secondary rates over a Gaussian Multiple Access Channel (MAC) has been considered under the assumption of an opportunistic interference cancellation (OIC) at the secondary receiver. In [22], system level capacity of a spectrum sharing communication network, under received average interference power constraints, has been studied. Therein, the capacity of two scenarios, namely cognitive Radio based central access network and cognitive Radio assisted virtual Multiple-Input Multiple-Output (MIMO) network, have been analyzed.

In order to generalize the existing results and to provide a uniform framework of performance limits of a spectrum sharing protocol under interference outage constraint, we analyze in this paper, the ergodic capacity under two different transmit-power constraints: a peak power constraint and an average power constraint. In each case, two different interference constraints at the primary user receiver are considered: interference outage constraint and a signal-to-interference (SI) outage constraint. The former outage events occur when the interference power at the primary user receiver is above a certain threshold, say $Q_{peak}$, whereas the later outage events happen when the ratio between the signal power and the interference power at the primary user receiver is below a certain threshold, say $I_{peak}$. Note that these constraints are necessary to ensure low error probability decoding at the primary user receiver at power-limited and interference-limited regimes, respectively. Furthermore, in our framework, we also assume that the secondary transmitter is only provided with imperfect ST-PR CSI. More specifically, our main contributions in this paper are as follows:

- Assuming that the ST is provided a noisy version of the ST-PR CSI, we introduce the instantaneous interference outage and signal-to-interference (SI) outage constraints that aim at protecting the primary user operating in a stringent delay-sensitive mode.
- Subject to both an average/peak power constraint and either an instantaneous interference outage or SI outage constraints, we derive the optimal power and the ergodic capacity of the secondary user operating in a spectrum sharing mode with the primary user, and highlight the effect of CSI error on the performance.
- We show that by letting the error variance of ST-PR CSI estimation tends toward one or zero, our framework extends naturally to no ST-PR CSI and perfect ST-PR CSI cases, and hence several previously reported results in the literature are retrieved as special cases.
- Specialized to Rayleigh fading channels, we provide asymptotic analysis of the derived results when the average or the peak power constraint tends to infinity.

The paper is organized as follows. Section II describes the system model. The optimal power profile and the ergodic capacity are derived according to an average and a peak transmit-power constraints and under different outage constraints, in Section III. Section IV addresses the perfect and no ST-PR CSI cases. In Section V, the derived results are applied to Rayleigh fading channels. Numerical results are briefly discussed in

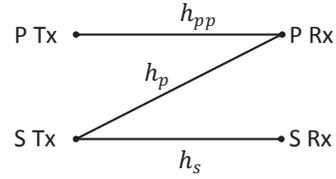

Fig. 1. A spectrum sharing channel model.

Section VI. Finally, Section VII concludes the paper.

*Notation:* The expectation operation is denoted by $\mathbb{E}\{\cdot\}$. The symbol $|x|$ is the modulus of the scalar $x$, while $[x]^+$ denotes $\max(0, x)$. The logarithms $\log(x)$ is the natural logarithm of $x$. A random variable is denoted by a bold face letter, e.g., $\mathbf{x}$, whereas the realization of $\mathbf{x}$ is denoted by $x$.

## II. System Model

We consider a spectrum sharing communication scenario as depicted in Fig.1, where a secondary user transmitter is communicating with a secondary user receiver, under certain constraints that will be defined later, through a licensed bandwidth occupied by a primary user. The signal received at the secondary user is given by:

$$\mathbf{r}_s(l) = \mathbf{h}_s(l)\,\mathbf{s}(l) + \mathbf{w}_s(l), \tag{1}$$

where $l$ is the discrete-time index, $\mathbf{s}(l)$ is the channel input, $\mathbf{h}_s(l)$ is the complex channel gain and $\mathbf{w}_s(l)$ is a zero-mean circularly symmetric complex white Gaussian noise with spectral density $N_0$ and is independent of $\mathbf{h}_s(l)$. The channel gains $\mathbf{h}_s(l)$, $\mathbf{h}_p(l)$ and $\mathbf{h}_{pp}(l)$ are assumed to be ergodic and stationary with continuous p.d.f. $f_{\mathbf{h}_s}(h_s)$, $f_{\mathbf{h}_p}(h_p)$ and $f_{\mathbf{h}_{pp}}(h_{pp})$, respectively. The secondary user transmitter is provided with the instantaneous CSI of the secondary user channel gain, $\mathbf{h}_s(l)$. However, it is only provided the statistics of $\mathbf{h}_p(l)$ through $f_{\mathbf{h}_p}(h_p)$ and a noisy version of $\mathbf{h}_p(l)$, say $\check{\mathbf{h}}_p(l)$, obtained via a band manager that coordinates the primary and the secondary users, or through a feedback link from the primary's receiver [6], [11], [16], [23]; such that $f_{\mathbf{h}_p|\check{\mathbf{h}}_p}(h_p|\check{h}_p)$ is also known. In order to improve its instantaneous estimate of $\mathbf{h}_p(l)$, the secondary transmitter further performs minimum mean square error (MMSE) estimation to obtain $\hat{\mathbf{h}}_p(l) = \mathbb{E}\left[\mathbf{h}_p(l)|\check{\mathbf{h}}_p(l) = \check{h}_p(l)\right]$. Note that to compute the MMSE estimate, the secondary transmitter needs to know the conditional p.d.f. of $\mathbf{h}_p(l)$ given $\check{\mathbf{h}}_p(l)$, which it does. Therefore, the ST-PR channel estimation model can be written as:

$$\mathbf{h}_p(l) = \sqrt{1-\sigma_p^2}\,\hat{\mathbf{h}}_p(l) + \sqrt{\sigma_p^2}\,\tilde{\mathbf{h}}_p(l), \tag{2}$$

where $\tilde{\mathbf{h}}_p(l)$ is the zero-mean unit-variance MMSE channel estimation error and $\sigma_p^2$ is the MMSE error variance. By well-known properties of the conditional mean, $\hat{\mathbf{h}}_p(l)$ and $\tilde{\mathbf{h}}_p(l)$ are uncorrelated. The channel estimation model (2) has been widely used in the channel estimation literature, e.g. [24], and recently in a CR context, e.g., [16], [18]. Furthermore, since the channel and the estimation models defined in (1) and (2), respectively, are stationary and memoryless, the capacity achieving statistics of the input $s(l)$ are also memoryless, independent and identically distributed (i.i.d.). Therefore, for simplicity we may drop the time index $l$ in (1) and (2). A sufficient statistic (including a noise variance normalization) to detect $\mathbf{s}$ from $\mathbf{r}_s$ in (1) is $\mathbf{y}_s = \frac{1}{\sqrt{N_0}}\left(\frac{\mathbf{h}_s^*}{|\mathbf{h}_s|}\mathbf{r}_s\right)$. The sufficient statistic $\mathbf{y}_s$ can be expressed by:

$$\mathbf{y}_s = |\mathbf{h}_s|\mathbf{x} + \mathbf{z}_s, \tag{3}$$

where $\mathbf{x} = \frac{1}{\sqrt{N_0}}\mathbf{s}$ and $\mathbf{z}_s = \frac{1}{\sqrt{N_0}}\left(\frac{\mathbf{h}_s^*}{|\mathbf{h}_s|}\mathbf{w}_s\right)$ is a zero-mean unit-variance white Gaussian noise. Let $P = \mathbb{E}[\mathbf{x}^2] = \frac{1}{N_0}\mathbb{E}[|\mathbf{s}|^2]$ be the normalized power at the secondary user transmitter. Since the sufficient statistic preserves the channel mutual information [25, Chap. 2], (3) does not entail any performance loss from a capacity point of view.

## III. Ergodic Capacity

For the channel given by (3), the ergodic capacity in nats per channel use (npcu), with transmitter and receiver side information and under either an average or a peak transmit-power constraint, can be expressed by [26]–[28]:

$$C = \max_P \mathbb{E}_{\mathbf{h}_s}\left[\ln\left(1 + P \cdot |\mathbf{h}_s|^2\right)\right]. \tag{4}$$

This is achievable using a variable-rate variable-power Gaussian codebook as described in [26], [28], or a simpler single Gaussian codebook with dynamic power allocation as argued in [27]. If no power adaptation is used in (4), then since the channel is i.i.d., the ergodic capacity with perfect CSI at the transmitter is equal to the one where only perfect CSI at the receiver is available. Therefore, to get the benefit of channel knowledge at the transmitter, it is necessary to allow the power $P$ to vary with the fading realizations, $h_s$. If additional constraints are to be considered, the ergodic capacity takes the form of the maximization (4) subject to these constraints. In particular, derivation of the secondary link ergodic capacity, in a cognitive radio setting, is subject to certain constraints related to the primary receiver that depend on the ST-PR link $h_p$. Since the secondary transmitter is provided an estimate of this channel gain, $\hat{h}_p$, then, it is natural to let the power also varies with $\hat{h}_p$, i.e., $P = P(h_s, \hat{h}_p)$.

### A. Average Transmit-Power And Interference Outage Constraints

In this subsection, the transmit power $P$ is subject to an average constraint: $\mathbb{E}\left[P(|\mathbf{h}_s|, |\hat{\mathbf{h}}_p|)\right] \leq P_{avg}$, where the expectation is over both $\mathbf{h}_s$ and $\hat{\mathbf{h}}_p$. Moreover, as the instantaneous CSI to the primary receiver is not available at the secondary transmitter, the probability that the interference power at the primary receiver be below a given positive threshold is always not null. However, we may resolve to tolerate a certain interference outage level and compute the capacity link consequently. Clearly, characterizing the capacity in terms of the interference outage constraint may be seen as a capacity-outage tradeoff: If no interference outage at the primary user is to be tolerated, the capacity of the secondary user link is equal to zero; on the other hand, if a high interference outage is acceptable, the capacity of the secondary user link is equal to the capacity as there is no-interference constraint. The ergodic capacity can be derived by solving the optimization problem:

$$C = \max_{P(\mathbf{h}_s, \hat{\mathbf{h}}_p)} \mathbb{E}_{\mathbf{h}_s, \hat{\mathbf{h}}_p}\left[\ln\left(1 + P(\mathbf{h}_s, \hat{\mathbf{h}}_p) \cdot |\mathbf{h}_s|^2\right)\right] \tag{5}$$

$$\text{s.t.} \quad \mathbb{E}_{\mathbf{h}_s, \hat{\mathbf{h}}_p}\left[P(\mathbf{h}_s, \hat{\mathbf{h}}_p)\right] \leq P_{avg} \tag{6}$$

$$\text{and } \text{Prob}\left\{P(\mathbf{h}_s, \hat{\mathbf{h}}_p) \cdot |\mathbf{h}_p|^2 \geq Q_{peak} \mid \mathbf{h}_s = h_s, \hat{\mathbf{h}}_p = \hat{h}_p\right\} \leq \epsilon. \tag{7}$$

Differently from [16] and [17], constraint (7) aims at reducing the instantaneous (not the average) interference power at the primary receiver for for all instantaneous values $h_s$ and $\hat{h}_p$. For simplicity, we will assume that $\mathbf{h}_p$ and $\mathbf{h}_s$ are independent, so that (7) is equivalent to:

$$P(h_s, \hat{h}_p) \leq \frac{Q_{peak}}{F^{-1}_{|\mathbf{h}_p|^2|\hat{\mathbf{h}}_p}(1-\epsilon)}, \tag{8}$$

where $F^{-1}_{|\mathbf{h}_p|^2|\hat{\mathbf{h}}_p}(\cdot)$ is the inverse cumulative distribution function (c.d.f.) of $|\mathbf{h}_p|^2$ conditioned on $\hat{\mathbf{h}}_p$. In order for $F^{-1}_{|\mathbf{h}_p|^2|\hat{\mathbf{h}}_p}(\cdot)$ to exist, it is sufficient that $f_{|\mathbf{h}_p|^2|\hat{\mathbf{h}}_p}$ be continuous and not null on an interval of its domain. Constraint (8) can be interpreted as a variable peak transmit-power constraint dictated by the interference constraint of the primary receiver. Therefore, the problem at hand is now equivalent to the derivation of the ergodic capacity under both variable peak and average transmit-power constraints. Recall that a somehow similar problem has been studied in [28] where the optimum power profile and the ergodic capacity have been partially found under both constant peak and average transmit-power constraints, but only in terms of Lagrangian multipliers. To provide a better understanding of the problem, an explicit solution in terms of system parameters is required. Furthermore, the peak constraint (8) is now depending on the ST-PR channel estimate $\hat{\mathbf{h}}_p$, thus, it is of interest to analyze the impact of such an estimation on cognitive radio performances. Indeed, using a similar approach than the one described in [13], [29] along with the Lagrangian method, it can be shown that the solution to the above convex optimization problem has the following water-filling power profile:

$$P(h_s, \hat{h}_p) = \min\left\{P_{|\mathbf{h}_p|^2|\hat{\mathbf{h}}_p}(\epsilon), \left[\frac{1}{\lambda} - \frac{1}{|h_s|^2}\right]^+\right\}, \tag{9}$$

where $\lambda$ is the positive Lagrange multiplier associated to constraint (6) and where $P_{|\mathbf{h}_p|^2|\hat{\mathbf{h}}_p}(\epsilon) = \frac{Q_{peak}}{F^{-1}_{|\mathbf{h}_p|^2|\hat{\mathbf{h}}_p}(1-\epsilon)}$ is the peak power constraint (8). Note that although the optimal power profile (9) has a similar form as the corresponding power profile of a dynamic time-division multiple-access (D-TDMA) in cognitive broadcast channel (C-BC) derived in [29, Theorem 4.1, Case II][1] and in [13, Theorem 2], our problem formulation is different from [13] and [29] in the following ways:

- Our framework deals with a spectrum sharing scenario where the secondary user has a noisy CSI of the cross link, and thus can capture the effect of such an uncertainty on the performance, whereas the formulation in [13] and [29], albeit more general, cannot encompass our setting.

---
[1] Although [29, Theorem 4.1, Case II] deals with a C-BC with $M$ primary users and $K$ secondary users, it is easy to see that by setting $M = K = 1$ therein, the power profile in [29, Theorem 4.1, Case II] coincides with (9).

- In our formulation, since the ST is only aware of a noisy version of the cross link CSI, then unlike [13] and [29], it cannot guarantee to limit the received interference power at PR in every cross link channel gain. Instead, the ST tries to opportunistically (using the cross link channel gain estimation $\hat{h}p$) protect the PR statistically by limiting the instantaneous outage probability at the primary receiver according to (7).
- In both [13] and [29], the Lagrange multiplier associated with the average power constraint, $\lambda$ ($g_{s1}$ in our manuscript), is found numerically, without a sufficient analytical insight. Note that finding $\lambda$ analytically (when possible) provides a better understanding as to how the power profile and hence the capacity depend on system parameters. Therefore, we give below an explicit solution of the optimal power (9) in terms of system parameters, and as such our formulation turns out o be more eloquent and insightful.

First, let us define the function $G(x)$ by:

$$G(x) = \frac{1 - F_{|\mathbf{h}_s|^2}(x)}{x} - \int_x^\infty \frac{f_{|\mathbf{h}_s|^2}(t)}{t} dt, \quad (10)$$

for $x > 0$, where $F_{|\mathbf{h}_s|^2}$ is the c.d.f. of $|\mathbf{h}_s|^2$. Since

$$\frac{f_{|\mathbf{h}_s|^2}(t)}{t} \leq \frac{f_{|\mathbf{h}_s|^2}(t)}{x},$$

for $t \geq x$ and $\int_x^\infty \frac{f_{|\mathbf{h}_s|^2}(t)}{x} dt$ exists, then so does $\int_x^\infty \frac{f_{|\mathbf{h}_s|^2}(t)}{t} dt$ and hence the function $G(\cdot)$ in (10) is well-defined. Since $G(x) < \frac{1}{x}$ and that $G(\cdot)$ is a decreasing continuous positive-definite function and thus invertible on $(0, \infty)$, then $G^{-1}(\cdot)$ is also a decreasing function on $(0, \infty)$. The purpose of defining such a function $G(\cdot)$ is to facilitate the presentation of the results in terms of system parameters. Clearly, for a system without cognition constraint, the optimal power is the well-known water-filling given by $P(h_s) = \left[\frac{1}{\lambda} - \frac{1}{|h_s|^2}\right]^+$, where $\lambda$ is obtained by solving the equation $\mathbb{E}_{|\mathbf{h}_s|^2}\left[\left[\frac{1}{\lambda} - \frac{1}{|h_s|^2}\right]^+\right] = P_{avg}$. The left hand side of the last equality is exactly $G(\lambda)$ and the definition in (10) follows from a simple integration by part. Note, for instance, that the definition in (10) and the properties of the function $G(\cdot)$ (continuous, monotonically decreasing, positive-definite on $(0, \infty)$) hold true for all class of fading channels considered in the paper. Therefore, the optimum power profile can be derived, with the help of the first order optimality conditions, as follows (please see Appendix for the proof):

- If $P_{avg} \geq \mathbb{E}_{\hat{\mathbf{h}}_p}\left[P_{|\mathbf{h}_p|^2|\hat{\mathbf{h}}_p}(\epsilon)\right]$, then we have:

$$P(h_s, \hat{h}_p) = P_{|\mathbf{h}_p|^2|\hat{\mathbf{h}}_p}(\epsilon) \quad (11)$$

- Otherwise, we have:
  - $P_{avg} > G(P_{|\mathbf{h}_p|^2|\hat{\mathbf{h}}_p}(\epsilon)^{-1})$

$$P(h_s, \hat{h}_p) = \begin{cases} 0 & |h_s|^2 < g_{s1} \\ \frac{1}{g_{s1}} - \frac{1}{|h_s|^2} & g_{s1} \leq |h_s|^2 < g_{s2} \\ P_{|\mathbf{h}_p|^2|\hat{\mathbf{h}}_p}(\epsilon) & |h_s|^2 \geq g_{s2} \end{cases} \quad (12)$$

  - $P_{avg} \leq G(P_{|\mathbf{h}_p|^2|\hat{\mathbf{h}}_p}(\epsilon)^{-1})$

$$P(h_s, \hat{h}_p) = \begin{cases} 0 & |h_s|^2 < g_{s1} \\ \frac{1}{g_{s1}} - \frac{1}{|h_s|^2} & |h_s|^2 \geq g_{s1}, \end{cases} \quad (13)$$

where $g_{s1}$ is obtained by satisfying the average power constraint (6) with equality, and where $g_{s2} = \left(\frac{1}{g_{s1}} - P_{|\mathbf{h}_p|^2|\hat{\mathbf{h}}_p}(\epsilon)\right)^{-1}$. In order to express $g_{s1}$ in terms of system parameters, let us define $S_x$ as a parametrized set that characterizes the values of $\hat{\mathbf{h}}_p$ which satisfy the inequality $F_{|\mathbf{h}_p|^2|\hat{\mathbf{h}}_p}(x Q_{peak}) < 1 - \epsilon$. Since the last inequality is equivalent to $x < \frac{1}{P_{|\mathbf{h}_p|^2|\hat{\mathbf{h}}_p}(\epsilon)}$, $S_x$ also characterizes the values of $\hat{\mathbf{h}}_p$ for which $\frac{1}{x} - P_{|\mathbf{h}_p|^2|\hat{\mathbf{h}}_p}(\epsilon) > 0$, i.e., $S_x = \left\{\hat{h}_p \mid x < \frac{1}{P_{|\mathbf{h}_p|^2|\hat{\mathbf{h}}_p}(\epsilon)}\right\}$. Note that substituting $x$ by $g_{s1}$, for instance, $S_{g_{s1}}$ would be the set of all $\hat{h}_p$ such that $g_{s2} > 0$ and hence the power profile is given by (12). Therefore, $g_{s1}$ can be expressed by (please see the Appendix for the derivation):

$$g_{s1} = K^{-1}(P_{avg}), \quad (14)$$

where $K(x)$ is defined on $(0, P_{|\mathbf{h}_p|^2|\hat{\mathbf{h}}_p}(\epsilon)]$ by:

$$K(x) = \begin{cases} G(x) - \mathbb{E}_{\hat{\mathbf{h}}_p \in S_x}\left[G\left(\left(1/x - P_{|\mathbf{h}_p|^2|\hat{\mathbf{h}}_p}(\epsilon)\right)^{-1}\right)\right] \\ \quad \text{if} \quad x < 1/P_{|\mathbf{h}_p|^2|\hat{\mathbf{h}}_p}(\epsilon) \\ G\left(1/P_{|\mathbf{h}_p|^2|\hat{\mathbf{h}}_p}(\epsilon)\right) \quad \text{if} \quad x = 1/P_{|\mathbf{h}_p|^2|\hat{\mathbf{h}}_p}(\epsilon) \end{cases} \quad (15)$$

Note that $g_{s1}$ in (14) is to be understood as the result of applying the inverse function of $K(x)$ to $P_{avg}$. It can be easily verified that the function $K(x)$ is continuous, monotonically decreasing and invertible, which guarantee the existence of $g_{s1}$. It should be highlighted that solving (14) to derive the optimal power profile is much more convenient than running a numerical optimization for each value of $P_{avg}$. Hence, combining (4), (11), (12) and (13), the secondary link capacity under average transmit-power and interference outage constraints is given by (16) at the top of the page. Note that (11) and the corresponding capacity expression in (16) clearly suggest that at high-power regime, the power profile and hence the capacity are impacted by the cross-link CSI only, irrespective to the secondary CSI. Such an insight provides, for instance, useful design guidelines that cannot be gained straightforwardly from (9), which is another advantage of our explicit solution over previous works.

### B. Average Transmit-Power And SI Outage Constraints

In the problem formulation of subsection III-A, the secondary transmitter does the best it can to ensure as low interference as possible to the primary receiver. However, constraint (7) does not guarantee a low-outage performance of the primary link. Recall that the outage at the primary receiver is defined as [30, Chap. 10]:

$$P_{out} = \text{Prob}\left\{\frac{P_{pp}|\mathbf{h}_{pp}|^2}{P(h_s, \hat{\mathbf{h}}_p)|\mathbf{h}_p|^2} \leq \lambda_{th} \text{ or } P_{pp}|\mathbf{h}_{pp}|^2 \leq P_{th}\right\},$$

where $\lambda_{th}$ and $P_{th}$ are SI power and signal power thresholds, respectively; and $P_{pp}$ is the primary link transmit power. For a sake of simplification, $P_{pp}$ is assumed to be independent

$$C = \begin{cases} \mathop{\mathrm{E}}_{|\mathbf{h}_s|^2, \hat{\mathbf{h}}_p}\left[\ln\left(1 + P_{|\mathbf{h}_p|^2|\hat{\mathbf{h}}_p}(\epsilon)|\mathbf{h}_s|^2\right)\right] & P_{avg} \geq \mathop{\mathrm{E}}_{\hat{\mathbf{h}}_p}\left[P_{|\mathbf{h}_p|^2|\hat{\mathbf{h}}_p}(\epsilon)\right] \\ \int_{t \geq g_{s2}} \ln\left(\tfrac{t}{g_{s1}}\right) f_{|\mathbf{h}_s|^2}(t)dt - \int_{\hat{h}_p \in S_{g_{s1}}} \int_{t \geq g_{s2}} \left[\ln\left(\tfrac{t}{g_{s1}}\right) - \ln\left(1 + P_{|\mathbf{h}_p|^2|\hat{\mathbf{h}}_p}(\epsilon)t\right)\right] f_{|\mathbf{h}_s|^2}(t) f_{\hat{\mathbf{h}}_p}(\hat{h}_p) \, d\hat{h}_p dt & P_{avg} < \mathop{\mathrm{E}}_{\hat{\mathbf{h}}_p}\left[P_{|\mathbf{h}_p|^2|\hat{\mathbf{h}}_p}(\epsilon)\right] \end{cases}. \quad (16)$$

of $h_{pp}$. Should the primary link be in deep fade, there is no chance to convey any information reliably on the primary link. A more engaged way to prevent interference outage at the primary receiver, albeit requiring more CSI at the secondary transmitter, would be to set a constraint on the SI power ratio as follows:

$$\mathrm{Prob}\left\{\frac{P_{pp}|\mathbf{h}_{pp}|^2}{P(\mathbf{h}_s, \hat{\mathbf{h}}_p)|\mathbf{h}_p|^2} \leq \lambda_{th} \mid \mathbf{h}_s = h_s, \hat{\mathbf{h}}_p = \hat{h}_p\right\} \leq \epsilon. \quad (17)$$

Letting $\beta = \frac{|\mathbf{h}_p|^2}{|\mathbf{h}_{pp}|^2}$ and substituting $|\mathbf{h}_p|^2$ and $Q_{peak}$ by $\beta$ and $\frac{P_{pp}}{\lambda_{th}}$, respectively; it is easy to see that (17) is equivalent to (8). Therefore, the optimum power profile and the ergodic capacity are given by (11), (12), (13) and (16), respectively, using the previous substitution.

### C. Peak Transmit-Power And Interference Or SI Outage Constraints

When instead of the average transmit-power constraint (6), a peak power constraint is to be respected

$$P(h_s, \hat{h}_p) \leq P_{peak}, \quad (18)$$

then either with the interference outage constraint (7) or SI outage constraint (17), the optimum power profile consists of transmitting with the maximum power subject to two peak power constraints. That is, $P(h_s, \hat{h}_p)$ is given by:

$$P(h_s, \hat{h}_p) = \frac{1}{h_{th}^2} \quad (19)$$

where $h_{th}^2 = \left(\min\left(P_{peak}, P_{\mathbf{X}|\hat{\mathbf{h}}_p}(\epsilon)\right)\right)^{-1}$ and $\mathbf{X}$ is either equal to $|\mathbf{h}_p|^2$ in case of interference outage constraint or is equal to $\beta$ in case of SI outage Constraint. Furthermore, the ergodic capacity is equal to:

$$C = \int_{\hat{h}_p} \left[\int_{|h_s|^2=0}^{\infty} \ln\left(1 + \frac{t}{h_{th}^2}\right) f_{|\mathbf{h}_s|^2}(t)dt\right] f_{\hat{\mathbf{h}}_p}(\hat{h}_p) \, d\hat{h}_p. \quad (20)$$

## IV. PERFECT AND NO ST-PR CSI CASES

In this section, the optimum power profile and the ergodic capacity, in case of perfect and no ST-PR CSI at the secondary transmitter, are obtained as special cases by letting $\sigma_p^2$ in (2) tends towards 0 and 1, respectively.

### A. Perfect ST-PR CSI

- Average transmit-power constraint

Recall that this special case has been studied in [13], where the power profile has been derived, but in terms of a Lagrange multiplier. We show below that our framework also captures this special case and a more explicit solution is presented. Indeed, when the secondary transmitter is provided the perfect instantaneous ST-PR channel gain $h_p$ ($\mathbf{h}_p = \hat{\mathbf{h}}_p$), the interference outage in (7) is equal to zero ($\epsilon = 0$) and $P_{|\mathbf{h}_p|^2|\hat{\mathbf{h}}_p}(\epsilon)$ in constraint (8) reduces to:

$$P_{|\mathbf{h}_p|^2|\hat{\mathbf{h}}_p}(\epsilon) = \frac{Q_{peak}}{|\hat{h}_p|^2}. \quad (21)$$

If the SI outage constraint is to be fulfilled, then (17) may be equivalently expressed by:

$$P_{\beta|\hat{\mathbf{h}}_p}(\epsilon) = \frac{Q_{peak}}{|\hat{h}_p|^2} \cdot \frac{1}{F^{-1}_{\frac{1}{|\mathbf{h}_{pp}|^2}}(1-\epsilon)}, \quad (22)$$

where $Q_{peak} = P_{pp}/\lambda_{th}$. Using the above substitutions, The optimum power profile can consequently be obtained from (11), (12) and (13). Noticing that $P_{\mathbf{X}|\hat{\mathbf{h}}_p}(\epsilon)$ in (21) and (22) depend on $\hat{h}_p$ only through its norm, then, substituting $f_{\hat{\mathbf{h}}_p}$ by $f_{|\hat{\mathbf{h}}_p|^2}$, the ergodic capacity is obtained from (16), with $S_{g_{s1}} = [\hat{h}_p^0, \infty[$ and $\hat{h}_p^0 = g_{s1} \times Q_{peak}$ in case of interference outage constraint, or $\hat{h}_p^0 = \frac{g_{s1} \times Q_{peak}}{F^{-1}_{\frac{1}{|\mathbf{h}_{pp}|^2}}(1-\epsilon)}$ in case of SI outage constraint.

- Peak transmit-power constraint

Similarly to the previous case, the optimum power is given by (19), with $h_{th}^2$ computed using (21) or (22) for signal outage constraint or SI outage constraint, respectively. The ergodic capacity can be simplified from (20) and is given by:

$$C = \int_0^{\infty} \left[\int_0^{\infty} \ln\left(1 + \frac{t}{h_{th}^2}\right) f_{|\mathbf{h}_s|^2}(t)dt\right] f_{|\hat{\mathbf{h}}_p|^2}(|\hat{h}_p|^2) \, d|\hat{h}_p|^2. \quad (23)$$

### B. No ST-PR CSI

With no instantaneous ST-PR CSI provided ($\mathbf{h}_p = \tilde{\mathbf{h}}_p$), the secondary transmitter can still rely on the statistics of $\mathbf{h}_p$ (through the p.d.f. $f_{\mathbf{h}_p}(\cdot)$) in order to respect the interference constraints. Note that now, the transmit-power depends only on $h_s$, i.e., $P = P(h_s)$. In this case, the interference outage (8) and SI outage (17) become:

$$P(h_s) \leq P_{\mathbf{X}|\hat{\mathbf{h}}_p}(\epsilon), \quad (24)$$

where $P_{\mathbf{X}|\hat{\mathbf{h}}_p}(\epsilon) = \frac{Q_{peak}}{F^{-1}_{\mathbf{X}}(1-\epsilon)}$, with $\mathbf{X}$ is again either equal to $|\mathbf{h}_p|^2$ in case of interference outage constraint or is equal to $\beta$ in case of SI outage constraint. Note that in this case, $P_{\mathbf{X}|\hat{\mathbf{h}}_p}(\epsilon)$ in (24) takes a fixed value and does not depend on $\hat{\mathbf{h}}_p$ since no ST-PR CSI is assumed. That is, $P_{\mathbf{X}|\hat{\mathbf{h}}_p}(\epsilon) = P_{\mathbf{X}}(\epsilon)$

- Average transmit-power constraint

The optimum power profile can be deduced from (11), (12) and (13) by replacing $P_{|\mathbf{h}_p|^2|\hat{\mathbf{h}}_p}(\epsilon)$ by $P_{\mathbf{X}}(\epsilon)$. Note that since now the peak constraint $P_{\mathbf{X}}(\epsilon)$ is constant, then the optimal power is either given by (11), or is given by (12) in which case $g_{s1} = \left(G(x) - G\left((1/x - P_{\mathbf{X}}(\epsilon))^{-1}\right)\right)^{-1}(P_{avg})$, or is given by (13) for which $g_{s1} = G^{-1}(P_{avg})$. Furthermore, the ergodic

$$C = \begin{cases} \mathbb{E}_{|\hat{\mathbf{h}}_s|^2}\left[\ln\left(1 + P_\mathbf{X}(\epsilon)|h_s|^2\right)\right] & P_{avg} \geq P_\mathbf{X}(\epsilon) \\ \int_{g_{s1}}^{g_{s2}} \ln\left(\frac{t}{g_{s1}}\right) f_{|\mathbf{h}_s|^2}(t)\, dt + \int_{g_{s2}}^{\infty} \ln\left(1 + P_\mathbf{X}(\epsilon) t\right) f_{|\mathbf{h}_s|^2}(t)\, dt & P_\mathbf{X}(\epsilon) > P_{avg} \geq G(P_\mathbf{X}(\epsilon)^{-1}) \\ \int_{g_{s1}}^{\infty} \ln\left(\frac{t}{g_{s1}}\right) f_{|\mathbf{h}_s|^2}(t)\, dt & P_{avg} < G(P_\mathbf{X}(\epsilon)^{-1}). \end{cases} \quad (25)$$

capacity is obtained by averaging (5) over $\mathbf{h}_s$ and is given by (25) at the top of the page.

- Peak transmit-power constraint

The optimal power profile is given by (19), with $P_{\mathbf{X}|\hat{\mathbf{h}}_p}(\epsilon)$ defined as in (24). Furthermore, the ergodic capacity is equal to:

$$C = \int_0^\infty \ln\left(1 + \frac{t}{h_{th}^2}\right) f_{|\mathbf{h}_s|^2}(t)\, dt. \quad (26)$$

## V. Application To Rayleigh Fading Channels

In this section, we assume that the channel gains $\mathbf{h}_s$, $\mathbf{h}_p$ and $\mathbf{h}_{pp}$ are i.i.d. zero-mean unit-variance circularly symmetric complex Gaussian random variables. Therefore, their square magnitude is exponentially distributed, the p.d.f.'s of $\beta$, $\frac{1}{|\mathbf{h}_{pp}|^2}$, and these of $|\mathbf{h}_p|^2$ and $\beta$ conditioned on $\hat{\mathbf{h}}_p$ are defined for $t \geq 0$, respectively by:

$$f_\beta(t) = \frac{1}{(1+t)^2}, \quad (27)$$

$$f_{\frac{1}{|\mathbf{h}_{pp}|^2}}(t) = \begin{cases} \frac{1}{t^2} e^{-\frac{1}{t}} & t > 0 \\ 0 & \text{otherwise}, \end{cases} \quad (28)$$

$$f_{|\mathbf{h}_p|^2|\hat{\mathbf{h}}_p}(t) = \frac{1}{\sigma_p^2} e^{-\frac{t+(1-\sigma_p^2)|\hat{h}_p|^2}{\sigma_p^2}} I_0\left(2\sqrt{\frac{(1-\sigma_p^2)|\hat{h}_p|^2 t}{\sigma_p^4}}\right), \quad (29)$$

$$f_{\beta|\hat{\mathbf{h}}_p}(t) = \frac{\sigma_p^4 + \left(\sigma_p^2 - \sigma_p^2|\hat{h}_p|^2 + |\hat{h}_p|^2\right) t}{\left(\sigma_p^2 + t\right)^3} e^{-\frac{(1-\sigma_p^2)|\hat{h}_p|^2}{\sigma_p^2 + t}}, \quad (30)$$

where $I_0(\cdot)$ in (29) is the modified Bessel function of the first kind. Their c.d.f. can be obtained in closed forms with the help of [30, Chap. 4 & 10] (for the last two p.d.f.) and are respectively given by:

$$F_\beta(t) = \frac{t}{1+t}, \quad (31)$$

$$F_{\frac{1}{|\mathbf{h}_{pp}|^2}}(t) = \begin{cases} e^{-\frac{1}{t}} & t > 0 \\ 0 & \text{otherwise}, \end{cases} \quad (32)$$

$$F_{|\mathbf{h}_p|^2|\hat{\mathbf{h}}_p}(t) = 1 - Q_1\left(\sqrt{\frac{2(1-\sigma_p^2)|\hat{h}_p|^2}{\sigma_p^2}}, \sqrt{\frac{2t}{\sigma_p^2}}\right), \quad (33)$$

$$F_{\beta|\hat{\mathbf{h}}_p}(t) = \frac{t}{\sigma_p^2 + t} e^{-\frac{(1-\sigma_p^2)|\hat{h}_p|^2}{\sigma_p^2 + t}}, \quad (34)$$

where $Q_1(\alpha, \beta)$ in (33) is the first order Marcum Q-Function defined by [30, Chap. 4]:

$$Q_1(\alpha, \beta) = \int_\beta^\infty x e^{-\frac{x^2+\alpha^2}{2}} I_0(\alpha x)\, dx. \quad (35)$$

The inverses c.d.f. of (31) and (32) are straightforward, whereas that of (34) can be easily derived (after few manipulations) in terms of the principal branch of the LambertW

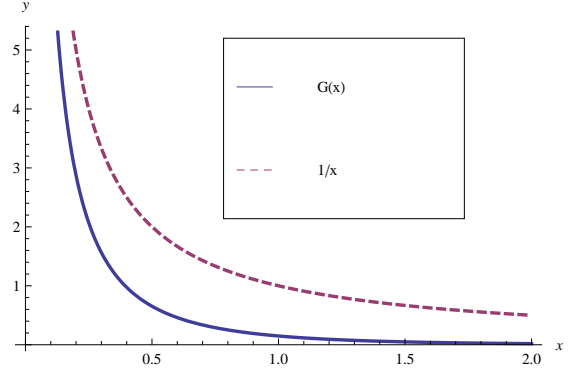

Fig. 2. A plot of $G(x)$ as defined in (10) for a Rayleigh fading secondary user channel: $G(x) = \frac{e^{-x}}{x} - \mathrm{E}_1(x)$, where $\mathrm{E}_1(x) = \int_x^\infty \frac{e^{-t}}{t} dt$ is the Exponential Integral Function [31, Chap. 5]

function, $W(.)$, as follows:

$$F_{\beta|\hat{\mathbf{h}}_p}^{-1}(u) = \frac{\sigma_p^2 W\left(\frac{(1-\sigma_p^2)|\hat{h}_p|^2}{\sigma_p^2} e^{-\frac{(1-\sigma_p^2)|\hat{h}_p|^2}{\sigma_p^2}} u\right)}{\frac{(1-\sigma_p^2)|\hat{h}_p|^2}{\sigma_p^2} - W\left(\frac{(1-\sigma_p^2)|\hat{h}_p|^2}{\sigma_p^2} e^{-\frac{(1-\sigma_p^2)|\hat{h}_p|^2}{\sigma_p^2}} u\right)}, \quad (36)$$

for $u \in [0, 1[$. Since $W$ is non-negative definite over $[0, \infty[$, then using the fact that $W(x \cdot e^x) = x$, it can be easily verified that $F_{\beta|\hat{\mathbf{h}}_p}^{-1}(u)$ is also non-negative definite over $u \in [0, 1[$. The inverse c.d.f. of (33) is unfortunately tedious to derive in a closed form, but can be computed numerically. Likewise, $g_{s1}$ can also be computed numerically using (14), where the function $G(\cdot)$, defined in (10), is now equal to:

$$G(x) = \frac{e^{-x}}{x} - \mathrm{E}_1(x),$$

where $\mathrm{E}_1(x) = \int_x^\infty \frac{e^{-t}}{t} dt$ is the Exponential Integral Function [31, Chap. 5]. A plot of $G(\cdot)$ is presented in Fig. 2.

### A. Results for $0 < \sigma_p^2 < 1$

The results regarding the optimum power profile and the ergodic capacity derived in Section III under average transmit-power and peak transmit-power are presented separately.

- Average transmit-power

The optimum power profile is directly obtained from (11), (12) and (13), using (33) (to compute $g_{s2}$ numerically) in case of interference outage constraint. Instead, substituting $|\mathbf{h}_p|^2$ and $Q_{peak}$ by $\beta$ and $\frac{P_{pp}}{\lambda_{th}}$, respectively, in (11), (12) and (13), and using (36), the optimum power profile is obtained in case of SI outage constraint. Since $F_{|\mathbf{h}_p|^2|\hat{\mathbf{h}}_p}$ and $F_{\beta|\hat{\mathbf{h}}_p}$ in (33) and (34) depend on $\hat{\mathbf{h}}_p$ only through its norm; and since they are

TABLE I
SUMMARY RESULTS OF THE OPTIMUM POWER PROFILE AND THE ERGODIC CAPACITY OF A SPECTRUM SHARING COGNITIVE RADIO NETWORK WHERE ALL THE CHANNELS ARE I.I.D. GAUSSIAN AND WITH NO ST-PR CSI ($\sigma_p^2 \to 1$).

| | Optimum Power Profile | given by (11), (12) and (13), with $P_{|\mathbf{h}_p|^2|\hat{h}_p}(\epsilon) = -\frac{Q_{peak}}{\ln(\epsilon)}$ |
|---|---|---|
| Average Transmit-Power And Interference Outage Constraints | Ergodic Capacity | $C = \begin{cases} \epsilon^{-1/Q_{peak}} \mathrm{E}_1\left(-\frac{\ln(\epsilon)}{Q_{peak}}\right) & P_{avg} \geq -\frac{Q_{peak}}{\ln(\epsilon)} \\ \mathrm{E}_1(g_{s1}) - \mathrm{E}_1(g_{s2}) + \epsilon^{-1/Q_{peak}} \mathrm{E}_1\left(\frac{g_{s2}^2}{g_{s2}-g_{s1}}\right) & -\frac{Q_{peak}}{\ln(\epsilon)} > P_{avg} \geq G\left(-\frac{\ln(\epsilon)}{Q_{peak}}\right) \\ \mathrm{E}_1(g_{s1}) & P_{avg} < G\left(-\frac{\ln(\epsilon)}{Q_{peak}}\right). \end{cases}$ |
| | Optimum Power Profile | given by (11), (12) and (13), with $P_{|\mathbf{h}_p|^2|\hat{h}_p}(\epsilon) = \frac{\epsilon}{1-\epsilon}Q_{peak}$ |
| Average Transmit-Power And SI Outage Constraints | Ergodic Capacity | $C = \begin{cases} e^{\frac{(1-\epsilon)/\epsilon}{Q_{peak}}} \mathrm{E}_1\left(\frac{(1-\epsilon)/\epsilon}{Q_{peak}}\right) & P_{avg} \geq \frac{\epsilon}{1-\epsilon}Q_{peak} \\ \mathrm{E}_1(g_{s1}) - \mathrm{E}_1(g_{s2}) + e^{\frac{(1-\epsilon)/\epsilon}{Q_{peak}}} \mathrm{E}_1\left(\frac{g_{s2}^2}{g_{s2}-g_{s1}}\right) & \frac{\epsilon}{1-\epsilon}Q_{peak} > P_{avg} \geq G\left(\frac{(1-\epsilon)/\epsilon}{Q_{peak}}\right) \\ \mathrm{E}_1(g_{s1}) & P_{avg} < G\left(\frac{(1-\epsilon)/\epsilon}{Q_{peak}}\right). \end{cases}$ |
| Peak Transmit-Power And Interference Outage Constraint | Optimum Power Profile | given by (19), with $h_{th}^2 = \left(\min\left(P_{peak}, -\frac{Q_{peak}}{\ln(\epsilon)}\right)\right)^{-1}$ |
| | Ergodic Capacity | $C = e^{h_{th}^2} \mathrm{E}_1\left(h_{th}^2\right)$ |
| Peak Transmit-Power And SI Outage Constraint | Optimum Power Profile | given by (19), with $h_{th}^2 = \left(\min\left(P_{peak}, \frac{\epsilon}{1-\epsilon}Q_{peak}\right)\right)^{-1}$ |
| | Ergodic Capacity | $C = e^{h_{th}^2} \mathrm{E}_1\left(h_{th}^2\right)$ |

both strictly decreasing functions in $|\hat{h}_p|^2$ for $0 < \sigma_p^2 < 1$ and for a given $t$, then substituting again $f_{\hat{\mathbf{h}}_p}$ by $f_{|\hat{\mathbf{h}}_p|^2}$, the ergodic capacity is obtained from (16), with $\mathcal{S}_{g_{s1}} = [[\hat{h}_p^0]^+, \infty[$, where $\hat{h}_p^0$ is the unique solution of $F_{\mathbf{X}|\hat{\mathbf{h}}_p}(g_{s1} \times Q_{peak}) = 1 - \epsilon$ and $\mathbf{X} = |\mathbf{h}_p|^2$ in case of interference outage constraint, or $\mathbf{X} = \beta$ in case of SI outage constraint. Nevertheless, in both cases, there is no closed form of the second integral in the capacity expression (16); albeit, a numerical evaluation is easy to obtain.

- Peak transmit-power

The power profile for interference outage and SI outage constraints is given by (19). Here again, the outer integral in the capacity expression (20), which also can be further simplified as in (23), is not easy to find analytically, and shall be computed numerically.

### B. Results For No ST-PR CSI Case

When no ST-PR CSI is available at the secondary transmitter ($\sigma_p^2 \to 1$), the capacity expression in (25) and (26) can interestingly be computed in closed forms. For convenience, the results regarding no ST-PR CSI case are presented in Table I. In this case, it is also of interest to point out the following facts:

- Average transmit-power constraint

From (11), (12) and (13), it can be seen that when $P_{avg} \to \infty$, then $P(h_s) \to P_{\mathbf{X}|\hat{\mathbf{h}}_p}(\epsilon)$, where $\mathbf{X} = \mathbf{h}_p$ in case of interference outage constraint, and $\mathbf{X} = \beta$ in case of SI outage constraint. Consequently, the ergodic capacity is equal to:

$$\lim_{P_{avg} \to \infty} C = e^{\frac{1}{P_{\mathbf{X}|\hat{\mathbf{h}}_p}(\epsilon)}} \mathrm{E}_1\left(\frac{1}{P_{\mathbf{X}|\hat{\mathbf{h}}_p}(\epsilon)}\right), \quad (37)$$

in agreement with [32, eq. (34)]. On the other hand, when no interference constraint is considered, i.e., $P_{\mathbf{X}|\hat{\mathbf{h}}_p}(\epsilon) \to \infty$ (this can be obtained by letting $\epsilon \to 1$), the optimum power is given by (13) and the capacity is equal to

$$\lim_{P_{\mathbf{X}|\hat{\mathbf{h}}_p}(\epsilon) \to \infty} C = \mathrm{E}_1(g_{s1}) = \mathrm{E}_1\left(G^{-1}(P_{avg})\right), \quad (38)$$

in agreement with [26].

- Peak transmit-power constraint

First, note that the optimal power profile is constant regardless of the instantaneous CSI, $h_s$. Hence, even without secondary CSI, one would achieve the same ergodic capacity in this case. Furthermore, if $P_{\mathbf{X}|\hat{\mathbf{h}}_p}(\epsilon) > P_{peak}$, then the optimum power profile and the ergodic capacity are equal to those of a fading channel with a peak transmit-power and without interference constraint. Moreover, increasing the power above $P_{\mathbf{X}|\hat{\mathbf{h}}_p}(\epsilon)$ does not provide any capacity gain. Furthermore, the infinite peak transmit power and infinite interference outage constraints limits can be computed from (26) and are respectively given by:

$$\lim_{P_{peak} \to \infty} C = e^{1/P_{\mathbf{X}|\hat{\mathbf{h}}_p}(\epsilon)} \mathrm{E}_1\left(1/P_{\mathbf{X}|\hat{\mathbf{h}}_p}(\epsilon)\right) \quad (39)$$

$$\lim_{P_{\mathbf{X}|\hat{\mathbf{h}}_p}(\epsilon) \to \infty} C = e^{1/P_{peak}} \mathrm{E}_1\left(1/P_{peak}\right) \quad (40)$$

### C. Results For Perfect ST-PR CSI Case

- Average transmit-power

When the secondary transmitter is aware of the instantaneous ST-PR CSI ($\sigma_p^2 \to 0$), the power profile is similarly given by (11), (12) and (13), along with (21) or (22) for interference outage or SI outage constraints, respectively. The threshold $h_p^0$ in the capacity expression (16) can be computed explicitly in both cases and is equal to $\hat{h}_p^0 = g_{s1} \times Q_{peak}$ or $\hat{h}_p^0 = g_{s1} \times Q_{peak} \ln\left(\frac{1}{1-\epsilon}\right)$, respectively. However, in both cases, there is no closed form of the second integral in the capacity expression (16) which may be evaluated numerically. Nevertheless, asymptotic analysis when $P_{avg}$ (respectively $P_{peak}$) is sufficiently large (no budget constraint) or the outage constraint is not effective ($Q_{peak}$ is sufficiently high, for instance), are provided below:

$$\lim_{P_{avg}\to\infty} C = \frac{Q_{eq}\ln(Q_{eq})}{Q_{eq}-1}, \quad (41)$$

where $Q_{eq} = Q_{peak}$ in case of interference constraint and $Q_{eq} = Q_{peak}\ln(1/(1-\epsilon))$ in case of SI outage constraint. Whereas the capacity at infinite outage constraint is given by (38), and is independent of the channel estimation quality.

- Peak transmit-power

In this case, the power profile (19) and the ergodic capacity (20) can be obtained using (21) or (22) for interference outage or SI outage constraints, respectively. At high peak power constraint, it can be shown that

$$\lim_{P_{peak}\to\infty} C = \frac{Q_{eq}\ln(Q_{eq})}{Q_{eq}-1} \quad (42)$$

whereas, the high interference outage constraint ergodic capacity is again independent of the channel estimation quality and is equal to the corresponding one where no ST-PR CSI is available (c.f. Table I).

## VI. NUMERICAL RESULTS

In this section, numerical results are provided for Rayleigh fading channels as derived in Section V. First come our results for an average transmit-power constraint. Figure 3 depicts the optimum power profile in function of the secondary channel $|h_s|^2$ for a given $\hat{h}_p$ value. As shown in Fig. 3, the power profile has typically two or three regions depending on the interference outage constraint value, $P_{\mathbf{X}|\hat{\mathbf{h}}_p}(\epsilon)$. For a relatively high $P_{\mathbf{X}|\hat{\mathbf{h}}_p}(\epsilon)$, the power profile is similar to a water-filling as discussed in [26]; whereas, a low $P_{\mathbf{X}|\hat{\mathbf{h}}_p}(\epsilon)$ value limits the power profile even when the secondary link, $|h_s|^2$ is "good". This limitation behavior affects also the ergodic capacity, no matter how the channel estimation quality is, as shown in Fig. 4. However, at low-power regime ($P_{avg} \to 0$), the ergodic capacity is insensitive to the interference outage constraint and equal capacity is achieved regardless of $P_{\mathbf{X}|\hat{\mathbf{h}}_p}(\epsilon)$ and $\sigma_p^2$. The capacity at infinite interference outage constraint and at infinite average power constraint are also shown in Fig. 4 as performance limits. Interestingly, the first limit is achieved at low-power regime. For instance, when $Q_{peak} = 10$ and $\epsilon = \epsilon_1 = 4.2\%$, so that $\frac{Q_{peak}}{\ln(1/\epsilon_1)} = 5$ dB, then the ergodic capacity with and without interference constraint is the same for $P_{avg}$ below 2 dB, irrespectively to $\sigma_p^2$. The second limit is achieved at relatively high $P_{avg}$ values and increasingly with respect to the channel estimation quality. In Fig. 5, the capacity loss percentage defined as the ratio of the difference between the ergodic capacity under perfect cross link CSI ($\sigma_p^2 \to 0$) and the same capacity where only a noisy cross link CSI with error variance $\sigma_p^2$ is available at the ST, over the capacity under perfect cross link CSI, is plotted for different $P_{avg}$ values. Figure 5 confirms our previous observations: At low $P_{avg} = 0$ dB, the capacity loss is equal to zero and the interference constraint has no effect on the secondary performance; at $P_{avg} = 6$ dB, the cross link CSI uncertainty impacts the secondary capacity and the loss may be over 15% for a poor CSI quality ($\sigma_p^2 \geq 0.85$). However, it is interesting to note that an average CSI quality ($\sigma_p^2 \leq 0.5$) is enough to contain the capacity loss (less than 6%); On the contrary, at $P_{avg} = 10$ dB, the cross link CSI quality is detrimental since the capacity loss may reach up to 40% and one requires a "very good" CSI quality to reduce the capacity loss. For instance, even with $\sigma_p^2 = 0.1$, the capacity loss is more than 10%. This is in fact expected since at high power regime, the capacity is dictated by the interference constraint only, as stipulated by (16).

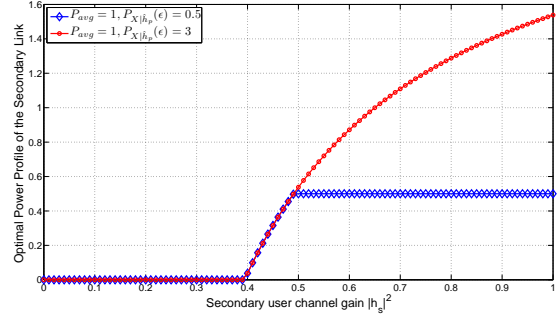

Fig. 3. Optimum power profile of the secondary link for a given $\hat{h}_p$ and under average transmit-power constraint, when $P_{avg} \leq \mathop{\mathrm{E}}\limits_{\hat{\mathbf{h}}_p}\left[P_{|\mathbf{h}_p|^2|\hat{h}_p}(\epsilon)\right]$.

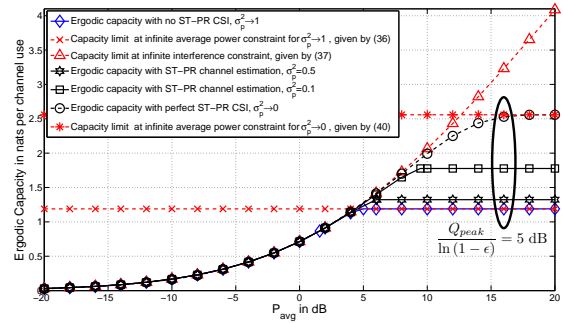

Fig. 4. Ergodic capacity of the secondary link under average transmit-power and interference outage constraints, for different channel estimation error variance values, $\sigma_p^2$.

In order to display results for SI outage constraint, we set $\frac{P_{pp}}{\lambda_{th}} = Q_{peak} = 10$ and $\epsilon = \epsilon_2 = 24\%$ in Fig. 6, so that the interference outage and SI outage constraints are equal in the no ST-PR CSI case. That is, $\frac{1}{\ln(1/\epsilon_1)}Q_{peak} = \frac{\epsilon_2}{1-\epsilon_2}Q_{peak} = 5$ dB, and hence the corresponding ergodic capacities of the secondary user under interference outage and SI outage constraints are equal. Note that to achieve equal ergodic capacity in this case, $\epsilon_2$ must be bigger than $\epsilon_1$ (a higher SI outage should be tolerated), which implies that at no ST-PR CSI case, SI outage constraint is more restrictive than interference outage constraint, from a capacity perspective, when $\frac{P_{pp}}{\lambda_{th}} = Q_{peak}$. Evidence of this can be seen by first noting that the $\epsilon_1 - \epsilon_2$ region $\mathcal{R}_1$ defined by:

$$\mathcal{R}_1 = \left\{(\epsilon_1, \epsilon_2) \mid \ln\frac{1}{\epsilon_1} \geq \frac{1}{\epsilon_2} - 1\right\}, \quad (43)$$

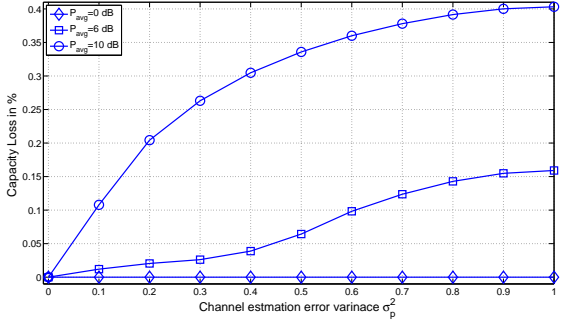

Fig. 5. Capacity Loss percentage of the secondary link under average transmit-power and interference outage constraints, versus channel estimation error variance $\sigma_p^2$, for different $P_{avg}$ values.

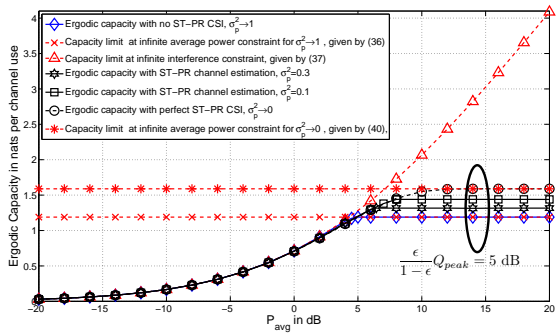

Fig. 6. Ergodic capacity of the secondary link under average transmit-power and SI outage constraints, for different channel estimation error variance values, $\sigma_p^2$.

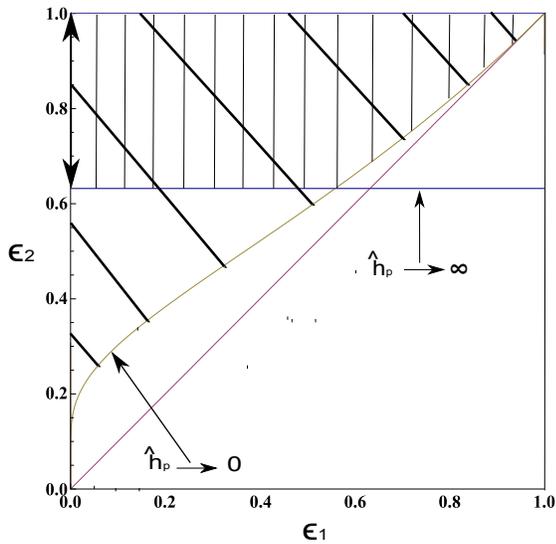

Fig. 7. $\epsilon_1 - \epsilon_2$ regions where the secondary link capacity under SI outage constraint is bigger than the one under interference outage for $\sigma_p^2 = 1$ in oblique lines, for $\sigma_p^2 = 0$ coinciding with y-axis such that $\epsilon_2 \geq 0.63$, and for $0 < \sigma_p^2 < 1$ in vertical lines.

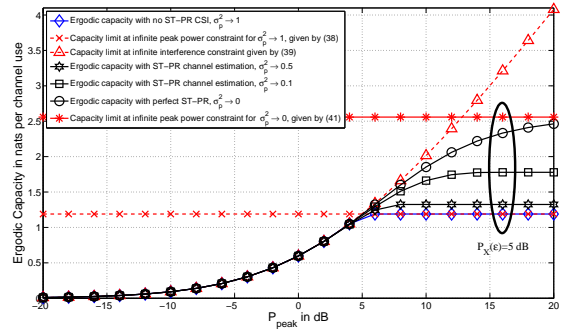

Fig. 8. Ergodic capacity of the secondary link under peak transmit-power constraint and for different channel estimation error variance values, $\sigma_p^2$.

corresponds to the region where the secondary user capacity under SI outage constraint $\epsilon_2$ is bigger than the one under interference outage constraint $\epsilon_1$, in the non ST-PR CSI scenario. Then using the inequality $x \geq \exp\left(1 - \frac{1}{x}\right)$, for all $x > 0$, (43) directly implies that $\epsilon_1 \leq \epsilon_2$. The fact that a higher SI outage is to be tolerated to achieve equal ergodic capacity than the one under interference outage constraint can also be seen for the perfect ST-PR CSI case, where $\epsilon_1 = 0$, by equalizing (21) and (22), which results in $\epsilon_1 - \epsilon_2$ region $\mathcal{R}_2$ defined by:

$$\mathcal{R}_2 = \{(\epsilon_1, \epsilon_2) \mid \epsilon_1 = 0, \epsilon_2 \geq 0.63\}, \quad (44)$$

and thus, all SI outage levels below this value, provide an ergodic capacity smaller than the one corresponding to an interference outage constraint. Only if the primary user is ready to accept an SI outage $\epsilon_2 \geq 63\%$, hence sacrificing its error probability performance, the secondary link would be able to achieve a higher ergodic capacity than the one under interference outage constraint. Furthermore, as shown in Fig. 6, even for such a high SI outage level ($\epsilon_2 = 24\%$), the secondary link ergodic capacity is smaller than the one displayed in Fig. 4 for equal channel estimation quality. When $0 < \sigma_p^2 < 1$, this fact can similarly be explained by noting that the secondary user capacity under SI outage is bigger than the one under interference outage iff $(\epsilon_1, \epsilon_2) \in \mathcal{R}_3$, where $\mathcal{R}_3$ is defined by:

$$\mathcal{R}_3 = \left\{(\epsilon_1, \epsilon_2) \mid P_{|\mathbf{h}_p|^2|\hat{\mathbf{h}}_p}(\epsilon_1) \leq P_{\beta|\hat{\mathbf{h}}_p}(\epsilon_2)\right\}, \quad (45)$$

for all $\hat{h}_p$. By letting $\hat{h}_p$ tends toward 0, it can be verified that $F^{-1}_{|\mathbf{h}_p|^2|\hat{\mathbf{h}}_p}(x) \to \sigma_p^2 \ln \frac{1}{x}$ and that $F^{-1}_{\beta|\hat{\mathbf{h}}_p}(x) \to \sigma_p^2 \ln \frac{x}{1-x}$. Hence, as $\hat{h}_p$ tends toward 0, (45) implies (43), which itself implies that $\epsilon_1 \leq \epsilon_2$ as proved above. Figure 7 illustrates the $\epsilon_1 - \epsilon_2$ regions $\mathcal{R}_1$, $\mathcal{R}_2$ and $\mathcal{R}_3$ defined by (43), (44) and (45), respectively.

Then come our results for a peak transmit-power constraint. Although the interference outage constraint limits the capacity at high-power regime, it has again no effect at low-power regime and no-interference performance is achieved as shown in Fig. 8. Finally, it is to be mentioned that at low power regime, the peak constraint is more stringent than the average constraint from a capacity point of view, in agreement with [33]. For instance, at a given interference outage constraint, say $P_X(\epsilon) = 5$ dB, the ergodic capacity provided by a peak transmit-power constraint $P_{peak} = 0$ dB is equal to 0.6 npcu, whereas for an average transmit-power $P_{avg} = 0$ dB, the ergodic capacity is equal to 0.71 npcu. This gap diminishes remarkably at high power regime where the outage constraint dictates the power profile.

## VII. Conclusion

A spectrum-sharing communication with ST-PR channel estimation at the secondary user has been addressed. The optimum power profile and the ergodic capacity have been derived for a class of fading channels with respect to an average or a peak transmit-power, along with more realistic interference outage constraints. The impact of channel estimation quality on the ergodic capacity has been highlighted. In all cases, asymptotic analysis has been discussed in order to provide a better understanding of the performance limits of a spectrum-sharing protocol. Our framework generalizes and encompasses several existent results.

## Appendix A
### Derivation of the optimum power profile given by (9) and equivalently by (11), (12) and (13)

To solve the optimization problem formulated by (5), (6) and (8), we first consider the set, say $\mathcal{A}$, of all $h_s$ and $\hat{h}_p$ values such that (8) is satisfied. That is, $\mathcal{A} = \left\{h_s, \hat{h}_p : P(h_s, \hat{h}_p) \leq P_{|\mathbf{h}_p|^2|\hat{h}_p}(\epsilon)\right\}$. Our optimization can thus be formulated by (5), (6), over the set $\mathcal{A}$. Since (5) is concave, (6) is linear and $\mathcal{A}$ is convex, then the new equivalent optimization problem is concave over $\mathcal{A}$. The solution of (5) and (6) over all channel realizations $h_s$ and $\hat{h}_p$ is known to be a water-filling given by $P(h_s, \hat{h}_p) = \left[\frac{1}{\lambda} - \frac{1}{|h_s|^2}\right]^+$. However, since we are now optimizing only over $\mathcal{A}$, the optimal power is capped by $P_{|\mathbf{h}_p|^2|\hat{h}_p}(\epsilon)$, which leads to (9).

Departing from (9), we show that the optimum power profile is given by (11), (12) and (13). First, note that since the optimization problem given by (5), (6) and (7) is convex with a feasible point, then $\lambda$ in (9) can be found by solving the (concave) Lagrange dual problem defined by [34]:

$$\min_{\lambda \geq 0} g(\lambda), \quad (46)$$

where $g(\lambda)$ is given by:

$$g(\lambda) = \mathop{\mathbb{E}}_{\mathbf{h}_s, \hat{\mathbf{h}}_p}\left[\ln\left(1 + P^*(\mathbf{h}_s, \hat{\mathbf{h}}_p) \cdot |\mathbf{h}_s|^2\right)\right] - \lambda\left(\mathop{\mathbb{E}}_{\mathbf{h}_s, \hat{\mathbf{h}}_p}\left[P^*(\mathbf{h}_s, \hat{\mathbf{h}}_p)\right] - P_{avg}\right)$$

and where $P^*(\mathbf{h}_s, \hat{\mathbf{h}}_p)$ is the optimal solution given by (9). Since $g(\lambda)$ is concave with a feasible minimum, then, we have:

- **C1**: If $\frac{\partial g}{\partial \lambda}|_{\lambda=0} \geq 0$, then $\min_{\lambda \geq 0} g(\lambda) = g(0)$,
- **C2**: otherwise, $\min_{\lambda \geq 0} g(\lambda) = g(g_{s1})$, for some $g_{s1} > 0$.

Now, it can be easily verified that

$$\frac{\partial g}{\partial \lambda} = \left(P_{avg} - \mathop{\mathbb{E}}_{\mathbf{h}_s, \hat{\mathbf{h}}_p}\left[P^*(\mathbf{h}_s, \hat{\mathbf{h}}_p)\right]\right),$$

and thus

$$\frac{\partial g}{\partial \lambda}|_{\lambda=0} = \left(P_{avg} - \mathop{\mathbb{E}}_{\hat{\mathbf{h}}_p}\left[P_{|\mathbf{h}_p|^2|\hat{h}_p}(\epsilon)\right]\right). \quad (47)$$

Combining (47), condition **C1** and condition **C2**, together with (9) yield (11), (12) and (13). It remains to prove (14). If **C2** holds, then by the complimentary slackness condition [34], the average power constraint is satisfied with equality, and we have:

$$P_{avg} = \mathop{\mathbb{E}}_{\mathbf{h}_s, \hat{\mathbf{h}}_p}\left[P^*(\mathbf{h}_s, \hat{\mathbf{h}}_p)\right]$$
$$= \mathop{\mathbb{E}}_{\hat{\mathbf{h}}_p}\left[\mathop{\mathbb{E}}_{\mathbf{h}_s|\hat{\mathbf{h}}_p}\left[P^*(\mathbf{h}_s, \hat{\mathbf{h}}_p) \mid \hat{\mathbf{h}}_p\right]\right]$$
$$= \mathop{\mathbb{E}}_{\hat{\mathbf{h}}_p \in \mathcal{S}_{g_{s1}}}\left[\mathop{\mathbb{E}}_{\mathbf{h}_s|\hat{\mathbf{h}}_p}\left[P^*(\mathbf{h}_s, \hat{\mathbf{h}}_p) \mid \hat{\mathbf{h}}_p \in \mathcal{S}_{g_{s1}}\right]\right]$$
$$+ \mathop{\mathbb{E}}_{\hat{\mathbf{h}}_p \in \bar{\mathcal{S}}_{g_{s1}}}\left[\mathop{\mathbb{E}}_{\mathbf{h}_s|\hat{\mathbf{h}}_p}\left[P^*(\mathbf{h}_s, \hat{\mathbf{h}}_p) \mid \hat{\mathbf{h}}_p \in \bar{\mathcal{S}}_{g_{s1}}\right]\right] \quad (48)$$

Now, the outer expectation in the second term on the right hand side of (48) is over all $\hat{\mathbf{h}}_p \in \bar{\mathcal{S}}_{g_{s1}}$ and hence the power profile is given by (13). Therefore, we can compute the second term in (48) as follows:

$$\mathop{\mathbb{E}}_{\hat{\mathbf{h}}_p \in \bar{\mathcal{S}}_{g_{s1}}}\left[\mathop{\mathbb{E}}_{\mathbf{h}_s|\hat{\mathbf{h}}_p}\left[P^*(\mathbf{h}_s, \hat{\mathbf{h}}_p) \mid \hat{\mathbf{h}}_p \in \bar{\mathcal{S}}_{g_{s1}}\right]\right] = \mathop{\mathbb{E}}_{\hat{\mathbf{h}}_p \in \bar{\mathcal{S}}_{g_{s1}}}\left[\mathop{\mathbb{E}}_{\mathbf{h}_s|\hat{\mathbf{h}}_p}\left[\frac{1}{g_{s1}} - \frac{1}{|h_s|^2}\right]^+\right]$$
$$= \mathop{\mathbb{E}}_{\hat{\mathbf{h}}_p \in \bar{\mathcal{S}}_{g_{s1}}}\left[\mathop{\mathbb{E}}_{\mathbf{h}_s}\left[\frac{1}{g_{s1}} - \frac{1}{|h_s|^2}\right]^+\right] \quad (49)$$
$$= \mathop{\mathbb{E}}_{\hat{\mathbf{h}}_p \in \bar{\mathcal{S}}_{g_{s1}}}\left[G(g_{s1})\right], \quad (50)$$

where (49) follows from the independence of $\mathbf{h}_s$ and $\hat{\mathbf{h}}_p$. On the contrary, the outer expectation in the first term on the right hand side of (48) is over all $\hat{\mathbf{h}}_p \in \mathcal{S}_{g_{s1}}$ and hence the power profile is given by (12). Therefore, we can compute the first term in (48) as shown at the bottom of the page. Now,

$$\mathop{\mathbb{E}}_{\hat{\mathbf{h}}_p \in \mathcal{S}_{g_{s1}}}\left[\mathop{\mathbb{E}}_{\mathbf{h}_s|\hat{\mathbf{h}}_p}\left[P^*(\mathbf{h}_s, \hat{\mathbf{h}}_p) \mid \hat{\mathbf{h}}_p \in \mathcal{S}_{g_{s1}}\right]\right] = \mathop{\mathbb{E}}_{\hat{\mathbf{h}}_p \in \mathcal{S}_{g_{s1}}}\left[\mathop{\mathbb{E}}_{|h_s|^2 \in [g_{s1}, g_{s2})}\left[\frac{1}{g_{s1}} - \frac{1}{|h_s|^2}\right]^+ + \mathop{\mathbb{E}}_{|h_s|^2 \geq g_{s2}}\left[P_{|\mathbf{h}_p|^2|\hat{h}_p}(\epsilon)\right]\right]$$
$$= \mathop{\mathbb{E}}_{\hat{\mathbf{h}}_p \in \mathcal{S}_{g_{s1}}}\left[G(g_{s1}) - \mathop{\mathbb{E}}_{|h_s|^2 > g_{s2}}\left[\frac{1}{g_{s1}} - \frac{1}{|h_s|^2} - P_{|\mathbf{h}_p|^2|\hat{h}_p}(\epsilon)\right]^+\right]$$
$$= \mathop{\mathbb{E}}_{\hat{\mathbf{h}}_p \in \mathcal{S}_{g_{s1}}}\left[G(g_{s1}) - G(g_{s2})\right],$$
$$= \mathop{\mathbb{E}}_{\hat{\mathbf{h}}_p \in \mathcal{S}_{g_{s1}}}\left[G(g_{s1})\right] - \mathop{\mathbb{E}}_{\hat{\mathbf{h}}_p \in \mathcal{S}_{g_{s1}}}\left[G(g_{s2})\right] \quad (51)$$

gathering (48), (50) and (51), we obtain:

$$P_{avg} = \mathbb{E}_{\hat{\mathbf{h}}_p}[G(g_{s1})] - \mathbb{E}_{\hat{\mathbf{h}}_p \in \mathcal{S}_{g_{s1}}}[G(g_{s2})]$$
$$= G(g_{s1}) - \mathbb{E}_{\hat{\mathbf{h}}_p \in \mathcal{S}_{g_{s1}}}[G(g_{s2})], \quad (52)$$

where (52) follows because $g_{s1}$ depends only on $P_{avg}$. Using the fact that the function $K(x)$ in (15) is invertible on $(0, P_{|\mathbf{h}_p|^2|\hat{h}_p}(\epsilon)]$, (14) follows immediately. Finally, since $K(x)$ is monotonically decreasing, then $g_{s1} < \frac{1}{P_{|\mathbf{h}_p|^2|\hat{h}_p}(\epsilon)}$ is equivalent to $P_{avg} > G\left(1/P_{|\mathbf{h}_p|^2|\hat{h}_p}(\epsilon)\right)$.

## Acknowledgment

The authors would like to thank Jose Roberto Ayala and Lokman Sboui for their constructive comments.

**Zouheir Rezki** (S'01, M'08) was born in Casablanca, Morocco. He received the Diplome d'Ingénieur degree from the École Nationale de l'Industrie Minérale (ENIM), Rabat, Morocco, in 1994, the M.Eng. degree from École de Technologie Supérieure, Montreal, Québec, Canada, in 2003, and the Ph.D. degree from École Polytechnique, Montreal, Québec, in 2008, all in electrical engineering. From October 2008 to September 2009, he was a postdoctoral research fellow with Data Communications Group, Department of Electrical and Computer Engineering, University of British Columbia. He is now a postdoctoral research fellow at King Abdullah University of Science and Technology (KAUST), Thuwal, Mekkah Province, Saudi Arabia. His research interests include: performance limits of communication systems, cognitive and sensor networks, physical-layer security, and low-complexity detection algorithms.

**Mohamed-Slim Alouini** (S'94, M'98, SM'03, F'09) was born in Tunis, Tunisia. He received the Ph.D. degree in electrical engineering from the California Institute of Technology (Caltech), Pasadena, CA, USA, in 1998. He was with the department of Electrical and Computer Engineering of the University of Minnesota, Minneapolis, MN, USA, then with the Electrical and Computer Engineering Program at the Texas A&M University at Qatar, Education City, Doha, Qatar. Since June 2009, he has been a Professor of Electrical Engineering in the Division of Physical Sciences and Engineering at KAUST, Saudi Arabia., where his current research interests include the design and performance analysis of wireless communication systems.

# Ergodic Capacity of Cognitive Radio under Imperfect Channel State Information

Zouheir Rezki *Member, IEEE,* and Mohamed-Slim Alouini, *Fellow, IEEE,*

*Abstract*—A spectrum-sharing communication system where the secondary user is aware of the instantaneous channel state information (CSI) of the secondary link, but knows only the statistics and an estimated version of the secondary transmitter-primary receiver (ST-PR) link, is investigated. The optimum power profile and the ergodic capacity of the secondary link are derived for general fading channels (with continuous probability density function) under average and peak transmit-power constraints and with respect to two different interference constraints: an interference outage constraint and a signal-to-interference outage constraint. When applied to Rayleigh fading channels, our results show, for instance, that the interference constraint is harmful at high-power regime in the sense that the capacity does not increase with the power, whereas at low-power regime, it has a marginal impact and no-interference performance corresponding to the ergodic capacity under average or peak transmit power constraint in absence of the primary user, may be achieved.

*Index Terms*—Cognitive radio, spectrum sharing, optimal power allocation, ergodic capacity and interference outage constraint.

## I. Introduction

Cognitive radio (CR) techniques have been proposed to efficiently use the spectrum through an adaptive, dynamic, and intelligent process [1]. Spectrum utilization can be improved by permitting a secondary user (who is not being serviced) to access a spectrum hole unoccupied by the primary user, or to share the spectrum with the primary user under certain interference constraints [2]. CR refers to different approaches to this problem that seek to overlay, underlay, or interweave the secondary user's signals with those of the primary users [3]. In the underlay settings, cognitive users can communicate as long as the interference caused to non cognitive users is below a certain threshold. Overlay systems, on the contrary, adopts a less conservative policy by permitting cognitive and non cognitive users to communicate simultaneously exploiting side information and using sophisticated coding techniques [4]. Perhaps the most conservative of the three, is the interweave system that permits to cognitive users to communicate provided that the actual spectrum is unoccupied by non cognitive users. More details on these three systems can be found, for instance, in [3], [4]. From an information-theoretical point-of-view, establishing performance limits of these systems relies strongly on the available side information that a cognitive user has about the network nodes: channel state information (CSI), coding techniques, codebooks, etc., e.g., [5]–[8].

In this paper, we focus on a spectrum-sharing CR model under general fading channels, with continuous probability density functions (p.d.f.), where the primary and the secondary users share the same spectrum under certain interference constraints. More specifically, we aim at analyzing the optimal power allocation and the ergodic capacity of the secondary link under limited channel knowledge at the secondary transmitter [9], [10].

Previous works have studied the impact of fading on the secondary link capacity under average or peak transmit-power, but assuming that the secondary transmitter is aware of the instantaneous CSI of the secondary transmitter-primary receiver (ST-PR) link, e.g., [11]–[14]. Although the later assumption generally guarantees an instantaneous limitation of the interference at the primary receiver, it is quite strong to obtain such valuable CSI in absence of an established cooperation protocol between the primary and the secondary links. Recall that protecting the primary user against interference may not be accurate if the CSI needed to estimate interference levels is coarsely precise, as shown in [15]. A step forward to address the problem in a more practical setting considering imperfect CSI has been realized in [16], [17], where the capacity or a lower bound on it has been derived under average received power or average interference outage constraint, respectively; but neither an average nor a peak transmit-power has been considered. The effect of ST-PR channel estimation at the SU on the ergodic capacity has also been analyzed under peak transmit power and peak interference constraint at the primary receiver, in [18]. However, unless some assumptions on the interference (strong or weak) caused by the primary user at the secondary receiver are adopted in the more general interference channel model therein, the results obtained seem to be an achievable rate using Gaussian codebook, as the capacity of the interference channel is still generally not known. Along similar lines, [19] considers the effect of statistical CSI rather than instantaneous channel estimation errors. Likewise, a rate-maximization problem of a secondary link where the cognitive users are equipped with multiple antennas and under an average transmit power along with an average interference constraints at the primary receiver has been considered in [20]. The secondary transmitter has been assumed to know the mean or the covariance of the ST-PR CSI through feedback, and in both cases, algorithms have been proposed to find the instantaneous optimum rate. In [21], a sum-rate maximization of the secondary rates over a Gaussian Multiple Access Channel (MAC) has been considered under the assumption of an opportunistic interference cancellation (OIC) at the secondary receiver. In [22], system level capacity of a spectrum sharing communication network, under received average interference power constraints, has been studied. Therein, the capacity of two scenarios, namely cognitive Radio based central access network and cognitive Radio assisted virtual Multiple-Input Multiple-Output (MIMO) network, have been analyzed.

In order to generalize the existing results and to provide a uniform framework of performance limits of a spectrum sharing protocol under interference outage constraint, we analyze in this paper, the ergodic capacity under two different transmit-power constraints: a peak power constraint and an average power constraint. In each case, two different interference constraints at the primary user receiver are considered: interference outage constraint and a signal-to-interference (SI) outage constraint. The former outage events occur when the interference power at the primary user receiver is above a certain threshold, say $Q_{peak}$, whereas the later outage events happen when the ratio between the signal power and the interference power at the primary user receiver is below a certain threshold, say $I_{peak}$. Note that these constraints are necessary to ensure low error probability decoding at the primary user receiver at power-limited and interference-limited regimes, respectively. Furthermore, in our framework, we also assume that the secondary transmitter is only provided with imperfect ST-PR CSI. More specifically, our main contributions in this paper are as follows:

- Assuming that the ST is provided a noisy version of the ST-PR CSI, we introduce the instantaneous interference outage and signal-to-interference (SI) outage constraints that aim at protecting the primary user operating in a stringent delay-sensitive mode.
- Subject to both an average/peak power constraint and either an instantaneous interference outage or SI outage constraints, we derive the optimal power and the ergodic capacity of the secondary user operating in a spectrum sharing mode with the primary user, and highlight the effect of CSI error on the performance.
- We show that by letting the error variance of ST-PR CSI estimation tends toward one or zero, our framework extends naturally to no ST-PR CSI and perfect ST-PR CSI cases, and hence several previously reported results in the literature are retrieved as special cases.
- Specialized to Rayleigh fading channels, we provide asymptotic analysis of the derived results when the average or the peak power constraint tends to infinity.

The paper is organized as follows. Section II describes the system model. The optimal power profile and the ergodic capacity are derived according to an average and a peak transmit-power constraints and under different outage constraints, in Section III. Section IV addresses the perfect and no ST-PR CSI cases. In Section V, the derived results are applied to Rayleigh fading channels. Numerical results are briefly discussed in Section VI. Finally, Section VII concludes the paper.

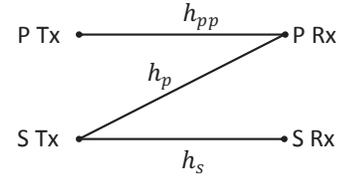

Fig. 1. A spectrum sharing channel model.

*Notation:* The expectation operation is denoted by $\mathbb{E}\{\cdot\}$. The symbol $|x|$ is the modulus of the scalar $x$, while $[x]^+$ denotes $\max(0, x)$. The logarithms $\log(x)$ is the natural logarithm of $x$. A random variable is denoted by a bold face letter, e.g., $\mathbf{x}$, whereas the realization of $\mathbf{x}$ is denoted by $x$.

## II. System Model

We consider a spectrum sharing communication scenario as depicted in Fig.1, where a secondary user transmitter is communicating with a secondary user receiver, under certain constraints that will be defined later, through a licensed bandwidth occupied by a primary user. The signal received at the secondary user is given by:

$$\mathbf{r}_s(l) = \mathbf{h}_s(l)\, \mathbf{s}(l) + \mathbf{w}_s(l), \tag{1}$$

where $l$ is the discrete-time index, $\mathbf{s}(l)$ is the channel input, $\mathbf{h}_s(l)$ is the complex channel gain and $\mathbf{w}_s(l)$ is a zero-mean circularly symmetric complex white Gaussian noise with spectral density $N_0$ and is independent of $\mathbf{h}_s(l)$. The channel gains $\mathbf{h}_s(l)$, $\mathbf{h}_p(l)$ and $\mathbf{h}_{pp}(l)$ are assumed to be ergodic and stationary with continuous p.d.f. $f_{\mathbf{h}_s}(h_s)$, $f_{\mathbf{h}_p}(h_p)$ and $f_{\mathbf{h}_{pp}}(h_{pp})$, respectively. The secondary user transmitter is provided with the instantaneous CSI of the secondary user channel gain, $\mathbf{h}_s(l)$. However, it is only provided the statistics of $\mathbf{h}_p(l)$ through $f_{\mathbf{h}_p}(h_p)$ and a noisy version of $\mathbf{h}_p(l)$, say $\check{\mathbf{h}}_p(l)$, obtained via a band manager that coordinates the primary and the secondary users, or through a feedback link from the primary's receiver [6], [11], [16], [23]; such that $f_{\mathbf{h}_p|\check{\mathbf{h}}_p}(h_p|\check{h}_p)$ is also known. In order to improve its instantaneous estimate of $\mathbf{h}_p(l)$, the secondary transmitter further performs minimum mean square error (MMSE) estimation to obtain $\hat{\mathbf{h}}_p(l) = \mathbb{E}\left[\mathbf{h}_p(l)|\check{\mathbf{h}}_p(l) = \check{h}_p(l)\right]$. Note that to compute the MMSE estimate, the secondary transmitter needs to know the conditional p.d.f. of $\mathbf{h}_p(l)$ given $\check{\mathbf{h}}_p(l)$, which it does. Therefore, the ST-PR channel estimation model can be written as:

$$\mathbf{h}_p(l) = \sqrt{1 - \sigma_p^2}\, \hat{\mathbf{h}}_p(l) + \sqrt{\sigma_p^2}\, \tilde{\mathbf{h}}_p(l), \tag{2}$$

where $\tilde{\mathbf{h}}_p(l)$ is the zero-mean unit-variance MMSE channel estimation error and $\sigma_p^2$ is the MMSE error variance. By well-known properties of the conditional mean, $\hat{\mathbf{h}}_p(l)$ and $\tilde{\mathbf{h}}_p(l)$ are uncorrelated. The channel estimation model (2) has been widely used in the channel estimation literature, e.g. [24], and recently in a CR context, e.g., [16], [18]. Furthermore, since the channel and the estimation models defined in (1) and (2), respectively, are stationary and memoryless, the capacity achieving statistics of the input $s(l)$ are also memoryless, independent and identically distributed (i.i.d.). Therefore, for simplicity we may drop the time index $l$ in (1) and (2). A sufficient statistic (including a noise variance normalization) to detect $\mathbf{s}$ from $\mathbf{r}_s$ in (1) is $\mathbf{y}_s = \frac{1}{\sqrt{N_0}} \left( \frac{\mathbf{h}_s^*}{|\mathbf{h}_s|} \mathbf{r}_s \right)$. The sufficient statistic $\mathbf{y}_s$ can be expressed by:

$$\mathbf{y}_s = |\mathbf{h}_s|\mathbf{x} + \mathbf{z}_s, \quad (3)$$

where $\mathbf{x} = \frac{1}{\sqrt{N_0}}\mathbf{s}$ and $\mathbf{z}_s = \frac{1}{\sqrt{N_0}}\left(\frac{\mathbf{h}_s^*}{|\mathbf{h}_s|}\mathbf{w}_s\right)$ is a zero-mean unit-variance white Gaussian noise. Let $P = \mathbb{E}[\mathbf{x}^2] = \frac{1}{N_0}\mathbb{E}[|\mathbf{s}|^2]$ be the normalized power at the secondary user transmitter. Since the sufficient statistic preserves the channel mutual information [25, Chap. 2], (3) does not entail any performance loss from a capacity point of view.

### III. Ergodic Capacity

For the channel given by (3), the ergodic capacity in nats per channel use (npcu), with transmitter and receiver side information and under either an average or a peak transmit-power constraint, can be expressed by [26]–[28]:

$$C = \max_P \mathbb{E}_{\mathbf{h}_s}\left[\ln\left(1 + P \cdot |\mathbf{h}_s|^2\right)\right]. \quad (4)$$

This is achievable using a variable-rate variable-power Gaussian codebook as described in [26], [28], or a simpler single Gaussian codebook with dynamic power allocation as argued in [27]. If no power adaptation is used in (4), then since the channel is i.i.d., the ergodic capacity with perfect CSI at the transmitter is equal to the one where only perfect CSI at the receiver is available. Therefore, to get the benefit of channel knowledge at the transmitter, it is necessary to allow the power $P$ to vary with the fading realizations, $h_s$. If additional constraints are to be considered, the ergodic capacity takes the form of the maximization (4) subject to these constraints. In particular, derivation of the secondary link ergodic capacity, in a cognitive radio setting, is subject to certain constraints related to the primary receiver that depend on the ST-PR link $h_p$. Since the secondary transmitter is provided an estimate of this channel gain, $\hat{h}_p$, then, it is natural to let the power also varies with $\hat{h}_p$, i.e., $P = P(h_s, \hat{h}_p)$.

#### A. Average Transmit-Power And Interference Outage Constraints

In this subsection, the transmit power $P$ is subject to an average constraint: $\mathbb{E}\left[P(|\mathbf{h}_s|, |\hat{\mathbf{h}}_p|)\right] \leq P_{avg}$, where the expectation is over both $\mathbf{h}_s$ and $\hat{\mathbf{h}}_p$. Moreover, as the instantaneous CSI to the primary receiver is not available at the secondary transmitter, the probability that the interference power at the primary receiver be below a given positive threshold is always not null. However, we may resolve to tolerate a certain interference outage level and compute the capacity link consequently. Clearly, characterizing the capacity in terms of the interference outage constraint may be seen as a capacity-outage tradeoff: If no interference outage at the primary user is to be tolerated, the capacity of the secondary user link is equal to zero; on the other hand, if a high interference outage is acceptable, the capacity of the secondary user link is equal to the capacity as there is no-interference constraint. The ergodic capacity can be derived by solving the optimization problem:

$$C = \max_{P(\mathbf{h}_s, \hat{\mathbf{h}}_p)} \mathbb{E}_{\mathbf{h}_s, \hat{\mathbf{h}}_p}\left[\ln\left(1 + P(\mathbf{h}_s, \hat{\mathbf{h}}_p) \cdot |\mathbf{h}_s|^2\right)\right] \quad (5)$$

$$\text{s.t.} \quad \mathbb{E}_{\mathbf{h}_s, \hat{\mathbf{h}}_p}\left[P(\mathbf{h}_s, \hat{\mathbf{h}}_p)\right] \leq P_{avg} \quad (6)$$

$$\text{and } \text{Prob}\left\{P(\mathbf{h}_s, \hat{\mathbf{h}}_p) \cdot |\mathbf{h}_p|^2 \geq Q_{peak} \mid \mathbf{h}_s = h_s, \hat{\mathbf{h}}_p = \hat{h}_p\right\} \leq \epsilon. \quad (7)$$

Differently from [16] and [17], constraint (7) aims at reducing the instantaneous (not the average) interference power at the primary receiver for for all instantaneous values $h_s$ and $\hat{h}_p$. For simplicity, we will assume that $\mathbf{h}_p$ and $\mathbf{h}_s$ are independent, so that (7) is equivalent to:

$$P(h_s, \hat{h}_p) \leq \frac{Q_{peak}}{F^{-1}_{|\mathbf{h}_p|^2|\hat{\mathbf{h}}_p}(1-\epsilon)}, \quad (8)$$

where $F^{-1}_{|\mathbf{h}_p|^2|\hat{\mathbf{h}}_p}(\cdot)$ is the inverse cumulative distribution function (c.d.f.) of $|\mathbf{h}_p|^2$ conditioned on $\hat{\mathbf{h}}_p$. In order for $F^{-1}_{|\mathbf{h}_p|^2|\hat{\mathbf{h}}_p}(\cdot)$ to exist, it is sufficient that $f_{|\mathbf{h}_p|^2|\hat{\mathbf{h}}_p}$ be continuous and not null on an interval of its domain. Constraint (8) can be interpreted as a variable peak transmit-power constraint dictated by the interference constraint of the primary receiver. Therefore, the problem at hand is now equivalent to the derivation of the ergodic capacity under both variable peak and average transmit-power constraints. Recall that a somehow similar problem has been studied in [28] where the optimum power profile and the ergodic capacity have been partially found under both constant peak and average transmit-power constraints, but only in terms of Lagrangian multipliers. To provide a better understanding of the problem, an explicit solution in terms of system parameters is required. Furthermore, the peak constraint (8) is now depending on the ST-PR channel estimate $\hat{\mathbf{h}}_p$, thus, it is of interest to analyze the impact of such an estimation on cognitive radio performances. Indeed, using a similar approach than the one described in [13], [29] along with the Lagrangian method, it can be shown that the solution to the above convex optimization problem has the following water-filling power profile:

$$P(h_s, \hat{h}_p) = \min\left\{P_{|\mathbf{h}_p|^2|\hat{\mathbf{h}}_p}(\epsilon), \left[\frac{1}{\lambda} - \frac{1}{|h_s|^2}\right]^+\right\}, \quad (9)$$

where $\lambda$ is the positive Lagrange multiplier associated to constraint (6) and where $P_{|\mathbf{h}_p|^2|\hat{\mathbf{h}}_p}(\epsilon) = \frac{Q_{peak}}{F^{-1}_{|\mathbf{h}_p|^2|\hat{\mathbf{h}}_p}(1-\epsilon)}$ is the peak power constraint (8). Note that although the optimal power profile (9) has a similar form as the corresponding power profile of a dynamic time-division multiple-access (D-TDMA) in cognitive broadcast channel (C-BC) derived in [29, Theorem 4.1, Case II][1] and in [13, Theorem 2], our problem formulation is different from [13] and [29] in the following ways:

- Our framework deals with a spectrum sharing scenario where the secondary user has a noisy CSI of the cross link, and thus can capture the effect of such an uncertainty

---

[1]Although [29, Theorem 4.1, Case II] deals with a C-BC with $M$ primary users and $K$ secondary users, it is easy to see that by setting $M = K = 1$ therein, the power profile in [29, Theorem 4.1, Case II] coincides with (9).

- on the performance, whereas the formulation in [13] and [29], albeit more general, cannot encompass our setting.
- In our formulation, since the ST is only aware of a noisy version of the cross link CSI, then unlike [13] and [29], it cannot guarantee to limit the received interference power at PR in every cross link channel gain. Instead, the ST tries to opportunistically (using the cross link channel gain estimation $\hat{h}p$) protect the PR statistically by limiting the instantaneous outage probability at the primary receiver according to (7).
- In both [13] and [29], the Lagrange multiplier associated with the average power constraint, $\lambda$ ($g_{s1}$ in our manuscript), is found numerically, without a sufficient analytical insight. Note that finding $\lambda$ analytically (when possible) provides a better understanding as to how the power profile and hence the capacity depend on system parameters. Therefore, we give below an explicit solution of the optimal power (9) in terms of system parameters, and as such our formulation turns out o be more eloquent and insightful.

First, let us define the function $G(x)$ by:

$$G(x) = \frac{1 - F_{|\mathbf{h}_s|^2}(x)}{x} - \int_x^\infty \frac{f_{|\mathbf{h}_s|^2}(t)}{t} dt, \quad (10)$$

for $x > 0$, where $F_{|\mathbf{h}_s|^2}$ is the c.d.f. of $|\mathbf{h}_s|^2$. Since

$$\frac{f_{|\mathbf{h}_s|^2}(t)}{t} \leq \frac{f_{|\mathbf{h}_s|^2}(t)}{x},$$

for $t \geq x$ and $\int_x^\infty \frac{f_{|\mathbf{h}_s|^2}(t)}{x} dt$ exists, then so does $\int_x^\infty \frac{f_{|\mathbf{h}_s|^2}(t)}{t} dt$ and hence the function $G(\cdot)$ in (10) is well-defined. Since $G(x) < \frac{1}{x}$ and that $G(\cdot)$ is a decreasing continuous positive-definite function and thus invertible on $(0, \infty)$, then $G^{-1}(\cdot)$ is also a decreasing function on $(0, \infty)$. The purpose of defining such a function $G(\cdot)$ is to facilitate the presentation of the results in terms of system parameters. Clearly, for a system without cognition constraint, the optimal power is the well-known water-filling given by $P(h_s) = \left[\frac{1}{\lambda} - \frac{1}{|h_s|^2}\right]^+$, where $\lambda$ is obtained by solving the equation $\mathbb{E}_{|\mathbf{h}_s|^2}\left[\left[\frac{1}{\lambda} - \frac{1}{|h_s|^2}\right]^+\right] = P_{avg}$. The left hand side of the last equality is exactly $G(\lambda)$ and the definition in (10) follows from a simple integration by part. Note, for instance, that the definition in (10) and the properties of the function $G(\cdot)$ (continuous, monotonically decreasing, positive-definite on $(0, \infty)$) hold true for all class of fading channels considered in the paper. Therefore, the optimum power profile can be derived, with the help of the first order optimality conditions, as follows (please see Appendix for the proof):

- If $P_{avg} \geq \mathbb{E}_{\hat{\mathbf{h}}_p}\left[P_{|\mathbf{h}_p|^2|\hat{\mathbf{h}}_p}(\epsilon)\right]$, then we have:

$$P(h_s, \hat{h}_p) = P_{|\mathbf{h}_p|^2|\hat{\mathbf{h}}_p}(\epsilon) \quad (11)$$

- Otherwise, we have:
    - $P_{avg} > G(P_{|\mathbf{h}_p|^2|\hat{\mathbf{h}}_p}(\epsilon)^{-1})$

$$P(h_s, \hat{h}_p) = \begin{cases} 0 & |h_s|^2 < g_{s1} \\ \frac{1}{g_{s1}} - \frac{1}{|h_s|^2} & g_{s1} \leq |h_s|^2 < g_{s2} \\ P_{|\mathbf{h}_p|^2|\hat{\mathbf{h}}_p}(\epsilon) & |h_s|^2 \geq g_{s2} \end{cases} \quad (12)$$

- $P_{avg} \leq G(P_{|\mathbf{h}_p|^2|\hat{\mathbf{h}}_p}(\epsilon)^{-1})$

$$P(h_s, \hat{h}_p) = \begin{cases} 0 & |h_s|^2 < g_{s1} \\ \frac{1}{g_{s1}} - \frac{1}{|h_s|^2} & |h_s|^2 \geq g_{s1}, \end{cases} \quad (13)$$

where $g_{s1}$ is obtained by satisfying the average power constraint (6) with equality, and where $g_{s2} = \left(\frac{1}{g_{s1}} - P_{|\mathbf{h}_p|^2|\hat{\mathbf{h}}_p}(\epsilon)\right)^{-1}$. In order to express $g_{s1}$ in terms of system parameters, let us define $S_x$ as a parametrized set that characterizes the values of $\hat{\mathbf{h}}_p$ which satisfy the inequality $F_{|\mathbf{h}_p|^2|\hat{\mathbf{h}}_p}(x\, Q_{peak}) < 1 - \epsilon$. Since the last inequality is equivalent to $x < \frac{1}{P_{|\mathbf{h}_p|^2|\hat{\mathbf{h}}_p}(\epsilon)}$, $S_x$ also characterizes the values of $\hat{\mathbf{h}}_p$ for which $\frac{1}{x} - P_{|\mathbf{h}_p|^2|\hat{\mathbf{h}}_p}(\epsilon) > 0$, i.e., $S_x = \left\{\hat{h}_p \,|\, x < \frac{1}{P_{|\mathbf{h}_p|^2|\hat{\mathbf{h}}_p}(\epsilon)}\right\}$. Note that substituting $x$ by $g_{s1}$, for instance, $S_{g_{s1}}$ would be the set of all $\hat{h}_p$ such that $g_{s2} > 0$ and hence the power profile is given by (12). Therefore, $g_{s1}$ can be expressed by (please see the Appendix for the derivation):

$$g_{s1} = K^{-1}(P_{avg}), \quad (14)$$

where $K(x)$ is defined on $(0, P_{|\mathbf{h}_p|^2|\hat{\mathbf{h}}_p}(\epsilon)]$ by:

$$K(x) = \begin{cases} G(x) - \mathop{\mathbb{E}}\limits_{\hat{\mathbf{h}}_p \in S_x}\left[G\left(\left(1/x - P_{|\mathbf{h}_p|^2|\hat{\mathbf{h}}_p}(\epsilon)\right)^{-1}\right)\right] \\ \qquad\qquad\qquad \text{if} \quad x < 1/P_{|\mathbf{h}_p|^2|\hat{\mathbf{h}}_p}(\epsilon) \\ G\left(1/P_{|\mathbf{h}_p|^2|\hat{\mathbf{h}}_p}(\epsilon)\right) \quad \text{if} \quad x = 1/P_{|\mathbf{h}_p|^2|\hat{\mathbf{h}}_p}(\epsilon) \end{cases} \quad (15)$$

Note that $g_{s1}$ in (14) is to be understood as the result of applying the inverse function of $K(x)$ to $P_{avg}$. It can be easily verified that the function $K(x)$ is continuous, monotonically decreasing and invertible, which guarantee the existence of $g_{s1}$. It should be highlighted that solving (14) to derive the optimal power profile is much more convenient than running a numerical optimization for each value of $P_{avg}$. Hence, combining (4), (11), (12) and (13), the secondary link capacity under average transmit-power and interference outage constraints is given by (16) at the top of the page. Note that (11) and the corresponding capacity expression in (16) clearly suggest that at high-power regime, the power profile and hence the capacity are impacted by the cross-link CSI only, irrespective to the secondary CSI. Such an insight provides, for instance, useful design guidelines that cannot be gained straightforwardly from (9), which is another advantage of our explicit solution over previous works.

### B. Average Transmit-Power And SI Outage Constraints

In the problem formulation of subsection III-A, the secondary transmitter does the best it can to ensure as low interference as possible to the primary receiver. However, constraint (7) does not guarantee a low-outage performance of the primary link. Recall that the outage at the primary receiver is defined as [30, Chap. 10]:

$$P_{out} = \text{Prob}\left\{\frac{P_{pp}|\mathbf{h}_{pp}|^2}{P(\mathbf{h}_s, \hat{\mathbf{h}}_p)|\mathbf{h}_p|^2} \leq \lambda_{th} \text{ or } P_{pp}|\mathbf{h}_{pp}|^2 \leq P_{th}\right\},$$

where $\lambda_{th}$ and $P_{th}$ are SI power and signal power thresholds, respectively; and $P_{pp}$ is the primary link transmit power. For a sake of simplification, $P_{pp}$ is assumed to be independent

$$C = \begin{cases} \mathop{\mathrm{E}}_{|\mathbf{h}_s|^2, \hat{\mathbf{h}}_p}\left[\ln\left(1 + P_{|\mathbf{h}_p|^2|\hat{\mathbf{h}}_p}(\epsilon)|\mathbf{h}_s|^2\right)\right] & P_{avg} \geq \mathop{\mathrm{E}}_{\hat{\mathbf{h}}_p}\left[P_{|\mathbf{h}_p|^2|\hat{\mathbf{h}}_p}(\epsilon)\right] \\ \int_{t \geq g_{s2}} \ln\left(\tfrac{t}{g_{s1}}\right) f_{|\mathbf{h}_s|^2}(t) dt - \int_{\hat{h}_p \in S_{g_{s1}}} \int_{t \geq g_{s2}} \left[\ln\left(\tfrac{t}{g_{s1}}\right) - \ln\left(1 + P_{|\mathbf{h}_p|^2|\hat{\mathbf{h}}_p}(\epsilon) t\right)\right] f_{|\mathbf{h}_s|^2}(t) f_{\hat{\mathbf{h}}_p}(\hat{h}_p) \, d\hat{h}_p dt & P_{avg} < \mathop{\mathrm{E}}_{\hat{\mathbf{h}}_p}\left[P_{|\mathbf{h}_p|^2|\hat{\mathbf{h}}_p}(\epsilon)\right] \end{cases}. \quad (16)$$

of $h_{pp}$. Should the primary link be in deep fade, there is no chance to convey any information reliably on the primary link. A more engaged way to prevent interference outage at the primary receiver, albeit requiring more CSI at the secondary transmitter, would be to set a constraint on the SI power ratio as follows:

$$\mathrm{Prob}\left\{ \frac{P_{pp}|\mathbf{h}_{pp}|^2}{P(\mathbf{h}_s, \hat{\mathbf{h}}_p)|\mathbf{h}_p|^2} \leq \lambda_{th} \mid \mathbf{h}_s = h_s, \hat{\mathbf{h}}_p = \hat{h}_p \right\} \leq \epsilon. \quad (17)$$

Letting $\beta = \frac{|\mathbf{h}_p|^2}{|\mathbf{h}_{pp}|^2}$ and substituting $|\mathbf{h}_p|^2$ and $Q_{peak}$ by $\beta$ and $\frac{P_{pp}}{\lambda_{th}}$, respectively; it is easy to see that (17) is equivalent to (8). Therefore, the optimum power profile and the ergodic capacity are given by (11), (12), (13) and (16), respectively, using the previous substitution.

### C. Peak Transmit-Power And Interference Or SI Outage Constraints

When instead of the average transmit-power constraint (6), a peak power constraint is to be respected

$$P(h_s, \hat{h}_p) \leq P_{peak}, \quad (18)$$

then either with the interference outage constraint (7) or SI outage constraint (17), the optimum power profile consists of transmitting with the maximum power subject to two peak power constraints. That is, $P(h_s, \hat{h}_p)$ is given by:

$$P(h_s, \hat{h}_p) = \frac{1}{h_{th}^2} \quad (19)$$

where $h_{th}^2 = \left(\min\left(P_{peak}, P_{\mathbf{X}|\hat{\mathbf{h}}_p}(\epsilon)\right)\right)^{-1}$ and $\mathbf{X}$ is either equal to $|\mathbf{h}_p|^2$ in case of interference outage constraint or is equal to $\beta$ in case of SI outage Constraint. Furthermore, the ergodic capacity is equal to:

$$C = \int_{\hat{h}_p} \left[ \int_{|h_s|^2=0}^{\infty} \ln\left(1 + \frac{t}{h_{th}^2}\right) f_{|\mathbf{h}_s|^2}(t) dt \right] f_{\hat{\mathbf{h}}_p}(\hat{h}_p) \, d\hat{h}_p. \quad (20)$$

## IV. PERFECT AND NO ST-PR CSI CASES

In this section, the optimum power profile and the ergodic capacity, in case of perfect and no ST-PR CSI at the secondary transmitter, are obtained as special cases by letting $\sigma_p^2$ in (2) tends towards 0 and 1, respectively.

### A. Perfect ST-PR CSI

- Average transmit-power constraint

Recall that this special case has been studied in [13], where the power profile has been derived, but in terms of a Lagrange multiplier. We show below that our framework also captures this special case and a more explicit solution is presented. Indeed, when the secondary transmitter is provided the perfect instantaneous ST-PR channel gain $h_p$ ($\mathbf{h}_p = \hat{\mathbf{h}}_p$), the interference outage in (7) is equal to zero ($\epsilon = 0$) and $P_{|\mathbf{h}_p|^2|\hat{\mathbf{h}}_p}(\epsilon)$ in constraint (8) reduces to:

$$P_{|\mathbf{h}_p|^2|\hat{\mathbf{h}}_p}(\epsilon) = \frac{Q_{peak}}{|\hat{h}_p|^2}. \quad (21)$$

If the SI outage constraint is to be fulfilled, then (17) may be equivalently expressed by:

$$P_{\beta|\hat{\mathbf{h}}_p}(\epsilon) = \frac{Q_{peak}}{|\hat{h}_p|^2} \cdot \frac{1}{F^{-1}_{\frac{1}{|\mathbf{h}_{pp}|^2}}(1-\epsilon)}, \quad (22)$$

where $Q_{peak} = P_{pp}/\lambda_{th}$. Using the above substitutions, The optimum power profile can consequently be obtained from (11), (12) and (13). Noticing that $P_{\mathbf{X}|\hat{\mathbf{h}}_p}(\epsilon)$ in (21) and (22) depend on $\hat{h}_p$ only through its norm, then, substituting $f_{\hat{\mathbf{h}}_p}$ by $f_{|\hat{\mathbf{h}}_p|^2}$, the ergodic capacity is obtained from (16), with $S_{g_{s1}} = [\hat{h}_p^0, \infty[$ and $\hat{h}_p^0 = g_{s1} \times Q_{peak}$ in case of interference outage constraint, or $\hat{h}_p^0 = \frac{g_{s1} \times Q_{peak}}{F^{-1}_{\frac{1}{|\mathbf{h}_{pp}|^2}}(1-\epsilon)}$ in case of SI outage constraint.

- Peak transmit-power constraint

Similarly to the previous case, the optimum power is given by (19), with $h_{th}^2$ computed using (21) or (22) for signal outage constraint or SI outage constraint, respectively. The ergodic capacity can be simplified from (20) and is given by:

$$C = \int_0^{\infty} \left[ \int_0^{\infty} \ln\left(1 + \frac{t}{h_{th}^2}\right) f_{|\mathbf{h}_s|^2}(t) dt \right] f_{|\hat{\mathbf{h}}_p|^2}(|\hat{h}_p|^2) \, d|\hat{h}_p|^2. \quad (23)$$

### B. No ST-PR CSI

With no instantaneous ST-PR CSI provided ($\mathbf{h}_p = \tilde{\mathbf{h}}_p$), the secondary transmitter can still rely on the statistics of $\mathbf{h}_p$ (through the p.d.f. $f_{\mathbf{h}_p}(\cdot)$) in order to respect the interference constraints. Note that now, the transmit-power depends only on $h_s$, i.e., $P = P(h_s)$. In this case, the interference outage (8) and SI outage (17) become:

$$P(h_s) \leq P_{\mathbf{X}|\hat{\mathbf{h}}_p}(\epsilon), \quad (24)$$

where $P_{\mathbf{X}|\hat{\mathbf{h}}_p}(\epsilon) = \frac{Q_{peak}}{F^{-1}_{\mathbf{X}}(1-\epsilon)}$, with $\mathbf{X}$ is again either equal to $|\mathbf{h}_p|^2$ in case of interference outage constraint or is equal to $\beta$ in case of SI outage constraint. Note that in this case, $P_{\mathbf{X}|\hat{\mathbf{h}}_p}(\epsilon)$ in (24) takes a fixed value and does not depend on $\hat{\mathbf{h}}_p$ since no ST-PR CSI is assumed. That is, $P_{\mathbf{X}|\hat{\mathbf{h}}_p}(\epsilon) = P_{\mathbf{X}}(\epsilon)$

- Average transmit-power constraint

The optimum power profile can be deduced from (11), (12) and (13) by replacing $P_{|\mathbf{h}_p|^2|\hat{\mathbf{h}}_p}(\epsilon)$ by $P_{\mathbf{X}}(\epsilon)$. Note that since now the peak constraint $P_{\mathbf{X}}(\epsilon)$ is constant, then the optimal power is either given by (11), or is given by (12) in which case $g_{s1} = \left(G(x) - G\left((1/x - P_{\mathbf{X}}(\epsilon))^{-1}\right)\right)^{-1}(P_{avg})$, or is given by (13) for which $g_{s1} = G^{-1}(P_{avg})$. Furthermore, the ergodic

$$C = \begin{cases} \mathbb{E}_{|\hat{\mathbf{h}}_s|^2}\left[\ln\left(1 + P_{\mathbf{X}}(\epsilon)|h_s|^2\right)\right] & P_{avg} \geq P_{\mathbf{X}}(\epsilon) \\ \int_{g_{s1}}^{g_{s2}} \ln\left(\frac{t}{g_{s1}}\right) f_{|\mathbf{h}_s|^2}(t)\,dt + \int_{g_{s2}}^{\infty} \ln\left(1 + P_{\mathbf{X}}(\epsilon) t\right) f_{|\mathbf{h}_s|^2}(t)\,dt & P_{\mathbf{X}}(\epsilon) > P_{avg} \geq G(P_{\mathbf{X}}(\epsilon)^{-1}) \\ \int_{g_{s1}}^{\infty} \ln\left(\frac{t}{g_{s1}}\right) f_{|\mathbf{h}_s|^2}(t)\,dt & P_{avg} < G(P_{\mathbf{X}}(\epsilon)^{-1}). \end{cases} \quad (25)$$

capacity is obtained by averaging (5) over $\mathbf{h}_s$ and is given by (25) at the top of the page.

- Peak transmit-power constraint

The optimal power profile is given by (19), with $P_{\mathbf{X}|\hat{\mathbf{h}}_p}(\epsilon)$ defined as in (24). Furthermore, the ergodic capacity is equal to:

$$C = \int_0^{\infty} \ln\left(1 + \frac{t}{h_{th}^2}\right) f_{|\mathbf{h}_s|^2}(t)\,dt. \quad (26)$$

## V. Application To Rayleigh Fading Channels

In this section, we assume that the channel gains $\mathbf{h}_s$, $\mathbf{h}_p$ and $\mathbf{h}_{pp}$ are i.i.d. zero-mean unit-variance circularly symmetric complex Gaussian random variables. Therefore, their square magnitude is exponentially distributed, the p.d.f.'s of $\beta$, $\frac{1}{|\mathbf{h}_{pp}|^2}$, and these of $|\mathbf{h}_p|^2$ and $\beta$ conditioned on $\hat{\mathbf{h}}_p$ are defined for $t \geq 0$, respectively by:

$$f_\beta(t) = \frac{1}{(1+t)^2}, \quad (27)$$

$$f_{\frac{1}{|\mathbf{h}_{pp}|^2}}(t) = \begin{cases} \frac{1}{t^2} e^{-\frac{1}{t}} & t > 0 \\ 0 & \text{otherwise}, \end{cases} \quad (28)$$

$$f_{|\mathbf{h}_p|^2|\hat{\mathbf{h}}_p}(t) = \frac{1}{\sigma_p^2} e^{-\frac{t + (1-\sigma_p^2)|\hat{h}_p|^2}{\sigma_p^2}} I_0\left(2\sqrt{\frac{(1-\sigma_p^2)|\hat{h}_p|^2 t}{\sigma_p^4}}\right), \quad (29)$$

$$f_{\beta|\hat{\mathbf{h}}_p}(t) = \frac{\sigma_p^4 + \left(\sigma_p^2 - \sigma_p^2|\hat{h}_p|^2 + |\hat{h}_p|^2\right)t}{\left(\sigma_p^2 + t\right)^3} e^{-\frac{(1-\sigma_p^2)|\hat{h}_p|^2}{\sigma_p^2 + t}}, \quad (30)$$

where $I_0(\cdot)$ in (29) is the modified Bessel function of the first kind. Their c.d.f. can be obtained in closed forms with the help of [30, Chap. 4 & 10] (for the last two p.d.f.) and are respectively given by:

$$F_\beta(t) = \frac{t}{1+t}, \quad (31)$$

$$F_{\frac{1}{|\mathbf{h}_{pp}|^2}}(t) = \begin{cases} e^{-\frac{1}{t}} & t > 0 \\ 0 & \text{otherwise}, \end{cases} \quad (32)$$

$$F_{|\mathbf{h}_p|^2|\hat{\mathbf{h}}_p}(t) = 1 - Q_1\left(\sqrt{\frac{2(1-\sigma_p^2)|\hat{h}_p|^2}{\sigma_p^2}}, \sqrt{\frac{2t}{\sigma_p^2}}\right), \quad (33)$$

$$F_{\beta|\hat{\mathbf{h}}_p}(t) = \frac{t}{\sigma_p^2 + t} e^{-\frac{(1-\sigma_p^2)|\hat{h}_p|^2}{\sigma_p^2 + t}}, \quad (34)$$

where $Q_1(\alpha, \beta)$ in (33) is the first order Marcum Q-Function defined by [30, Chap. 4]:

$$Q_1(\alpha, \beta) = \int_\beta^\infty x e^{-\frac{x^2 + \alpha^2}{2}} I_0(\alpha x)\,dx. \quad (35)$$

The inverses c.d.f. of (31) and (32) are straightforward, whereas that of (34) can be easily derived (after few manipulations) in terms of the principal branch of the LambertW

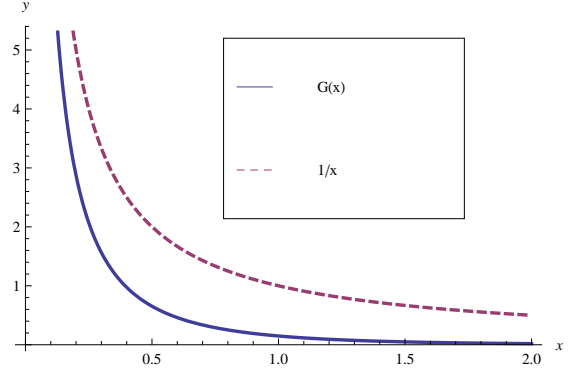

Fig. 2. A plot of $G(x)$ as defined in (10) for a Rayleigh fading secondary user channel: $G(x) = \frac{e^{-x}}{x} - E_1(x)$, where $E_1(x) = \int_x^\infty \frac{e^{-t}}{t} dt$ is the Exponential Integral Function [31, Chap. 5]

function, $W(.)$, as follows:

$$F_{\beta|\hat{\mathbf{h}}_p}^{-1}(u) = \frac{\sigma_p^2 W\left(\frac{(1-\sigma_p^2)|\hat{h}_p|^2}{\sigma_p^2} e^{-\frac{(1-\sigma_p^2)|\hat{h}_p|^2}{\sigma_p^2}} u\right)}{\frac{(1-\sigma_p^2)|\hat{h}_p|^2}{\sigma_p^2} - W\left(\frac{(1-\sigma_p^2)|\hat{h}_p|^2}{\sigma_p^2} e^{-\frac{(1-\sigma_p^2)|\hat{h}_p|^2}{\sigma_p^2}} u\right)}, \quad (36)$$

for $u \in [0, 1[$. Since $W$ is non-negative definite over $[0, \infty[$, then using the fact that $W(x \cdot e^x) = x$, it can be easily verified that $F_{\beta|\hat{\mathbf{h}}_p}^{-1}(u)$ is also non-negative definite over $u \in [0, 1[$. The inverse c.d.f. of (33) is unfortunately tedious to derive in a closed form, but can be computed numerically. Likewise, $g_{s1}$ can also be computed numerically using (14), where the function $G(\cdot)$, defined in (10), is now equal to:

$$G(x) = \frac{e^{-x}}{x} - E_1(x),$$

where $E_1(x) = \int_x^\infty \frac{e^{-t}}{t}\,dt$ is the Exponential Integral Function [31, Chap. 5]. A plot of $G(\cdot)$ is presented in Fig. 2.

### A. Results for $0 < \sigma_p^2 < 1$

The results regarding the optimum power profile and the ergodic capacity derived in Section III under average transmit-power and peak transmit-power are presented separately.

- Average transmit-power

The optimum power profile is directly obtained from (11), (12) and (13), using (33) (to compute $g_{s2}$ numerically) in case of interference outage constraint. Instead, substituting $|\mathbf{h}_p|^2$ and $Q_{peak}$ by $\beta$ and $\frac{P_{pp}}{\lambda_{th}}$, respectively, in (11), (12) and (13), and using (36), the optimum power profile is obtained in case of SI outage constraint. Since $F_{|\mathbf{h}_p|^2|\hat{\mathbf{h}}_p}$ and $F_{\beta|\hat{\mathbf{h}}_p}$ in (33) and (34) depend on $\hat{\mathbf{h}}_p$ only through its norm; and since they are

TABLE I
SUMMARY RESULTS OF THE OPTIMUM POWER PROFILE AND THE ERGODIC CAPACITY OF A SPECTRUM SHARING COGNITIVE RADIO NETWORK WHERE ALL THE CHANNELS ARE I.I.D. GAUSSIAN AND WITH NO ST-PR CSI ($\sigma_p^2 \to 1$).

| | Optimum Power Profile | given by (11), (12) and (13), with $P_{|\mathbf{h}_p|^2|\hat{h}_p}(\epsilon) = -\frac{Q_{peak}}{\ln(\epsilon)}$ |
|---|---|---|
| Average Transmit-Power And Interference Outage Constraints | Ergodic Capacity | $C = \begin{cases} \epsilon^{-1/Q_{peak}} \mathrm{E}_1\left(-\frac{\ln(\epsilon)}{Q_{peak}}\right) & P_{avg} \geq -\frac{Q_{peak}}{\ln(\epsilon)} \\ \mathrm{E}_1(g_{s1}) - \mathrm{E}_1(g_{s2}) + \epsilon^{-1/Q_{peak}} \mathrm{E}_1\left(\frac{g_{s2}^2}{g_{s2}-g_{s1}}\right) & -\frac{Q_{peak}}{\ln(\epsilon)} > P_{avg} \geq G\left(-\frac{\ln(\epsilon)}{Q_{peak}}\right) \\ \mathrm{E}_1(g_{s1}) & P_{avg} < G\left(-\frac{\ln(\epsilon)}{Q_{peak}}\right). \end{cases}$ |
| | Optimum Power Profile | given by (11), (12) and (13), with $P_{|\mathbf{h}_p|^2|\hat{h}_p}(\epsilon) = \frac{\epsilon}{1-\epsilon}Q_{peak}$ |
| Average Transmit-Power And SI Outage Constraints | Ergodic Capacity | $C = \begin{cases} e^{\frac{(1-\epsilon)/\epsilon}{Q_{peak}}} \mathrm{E}_1\left(\frac{(1-\epsilon)/\epsilon}{Q_{peak}}\right) & P_{avg} \geq \frac{\epsilon}{1-\epsilon}Q_{peak} \\ \mathrm{E}_1(g_{s1}) - \mathrm{E}_1(g_{s2}) + e^{\frac{(1-\epsilon)/\epsilon}{Q_{peak}}} \mathrm{E}_1\left(\frac{g_{s2}^2}{g_{s2}-g_{s1}}\right) & \frac{\epsilon}{1-\epsilon}Q_{peak} > P_{avg} \geq G\left(\frac{(1-\epsilon)/\epsilon}{Q_{peak}}\right) \\ \mathrm{E}_1(g_{s1}) & P_{avg} < G\left(\frac{(1-\epsilon)/\epsilon}{Q_{peak}}\right). \end{cases}$ |
| Peak Transmit-Power And Interference Outage Constraint | Optimum Power Profile | given by (19), with $h_{th}^2 = \left(\min\left(P_{peak}, -\frac{Q_{peak}}{\ln(\epsilon)}\right)\right)^{-1}$ |
| | Ergodic Capacity | $C = e^{h_{th}^2} \mathrm{E}_1\left(h_{th}^2\right)$ |
| Peak Transmit-Power And SI Outage Constraint | Optimum Power Profile | given by (19), with $h_{th}^2 = \left(\min\left(P_{peak}, \frac{\epsilon}{1-\epsilon}Q_{peak}\right)\right)^{-1}$ |
| | Ergodic Capacity | $C = e^{h_{th}^2} \mathrm{E}_1\left(h_{th}^2\right)$ |

both strictly decreasing functions in $|\hat{h}_p|^2$ for $0 < \sigma_p^2 < 1$ and for a given $t$, then substituting again $f_{\hat{\mathbf{h}}_p}$ by $f_{|\hat{\mathbf{h}}_p|^2}$, the ergodic capacity is obtained from (16), with $\mathcal{S}_{g_{s1}} = [[\hat{h}_p^0]^+, \infty[$, where $\hat{h}_p^0$ is the unique solution of $F_{\mathbf{X}|\hat{\mathbf{h}}_p}(g_{s1} \times Q_{peak}) = 1 - \epsilon$ and $\mathbf{X} = |\mathbf{h}_p|^2$ in case of interference outage constraint, or $\mathbf{X} = \beta$ in case of SI outage constraint. Nevertheless, in both cases, there is no closed form of the second integral in the capacity expression (16); albeit, a numerical evaluation is easy to obtain.

- Peak transmit-power

The power profile for interference outage and SI outage constraints is given by (19). Here again, the outer integral in the capacity expression (20), which also can be further simplified as in (23), is not easy to find analytically, and shall be computed numerically.

### B. Results For No ST-PR CSI Case

When no ST-PR CSI is available at the secondary transmitter ($\sigma_p^2 \to 1$), the capacity expression in (25) and (26) can interestingly be computed in closed forms. For convenience, the results regarding no ST-PR CSI case are presented in Table I. In this case, it is also of interest to point out the following facts:

- Average transmit-power constraint

From (11), (12) and (13), it can be seen that when $P_{avg} \to \infty$, then $P(h_s) \to P_{\mathbf{X}|\hat{\mathbf{h}}_p}(\epsilon)$, where $\mathbf{X} = \mathbf{h}_p$ in case of interference outage constraint, and $\mathbf{X} = \beta$ in case of SI outage constraint. Consequently, the ergodic capacity is equal to:

$$\lim_{P_{avg} \to \infty} C = e^{\frac{1}{P_{\mathbf{X}|\hat{\mathbf{h}}_p}(\epsilon)}} \mathrm{E}_1\left(\frac{1}{P_{\mathbf{X}|\hat{\mathbf{h}}_p}(\epsilon)}\right), \quad (37)$$

in agreement with [32, eq. (34)]. On the other hand, when no interference constraint is considered, i.e., $P_{\mathbf{X}|\hat{\mathbf{h}}_p}(\epsilon) \to \infty$ (this can be obtained by letting $\epsilon \to 1$), the optimum power is given by (13) and the capacity is equal to

$$\lim_{P_{\mathbf{X}|\hat{\mathbf{h}}_p}(\epsilon) \to \infty} C = \mathrm{E}_1(g_{s1}) = \mathrm{E}_1\left(G^{-1}(P_{avg})\right), \quad (38)$$

in agreement with [26].

- Peak transmit-power constraint

First, note that the optimal power profile is constant regardless of the instantaneous CSI, $h_s$. Hence, even without secondary CSI, one would achieve the same ergodic capacity in this case. Furthermore, if $P_{\mathbf{X}|\hat{\mathbf{h}}_p}(\epsilon) > P_{peak}$, then the optimum power profile and the ergodic capacity are equal to those of a fading channel with a peak transmit-power and without interference constraint. Moreover, increasing the power above $P_{\mathbf{X}|\hat{\mathbf{h}}_p}(\epsilon)$ does not provide any capacity gain. Furthermore, the infinite peak transmit power and infinite interference outage constraints limits can be computed from (26) and are respectively given by:

$$\lim_{P_{peak} \to \infty} C = e^{1/P_{\mathbf{X}|\hat{\mathbf{h}}_p}(\epsilon)} \mathrm{E}_1\left(1/P_{\mathbf{X}|\hat{\mathbf{h}}_p}(\epsilon)\right) \quad (39)$$

$$\lim_{P_{\mathbf{X}|\hat{\mathbf{h}}_p}(\epsilon) \to \infty} C = e^{1/P_{peak}} \mathrm{E}_1\left(1/P_{peak}\right) \quad (40)$$

### C. Results For Perfect ST-PR CSI Case

- Average transmit-power

When the secondary transmitter is aware of the instantaneous ST-PR CSI ($\sigma_p^2 \to 0$), the power profile is similarly given by (11), (12) and (13), along with (21) or (22) for interference outage or SI outage constraints, respectively. The threshold $\hat{h}_p^0$ in the capacity expression (16) can be computed explicitly in both cases and is equal to $\hat{h}_p^0 = g_{s1} \times Q_{peak}$ or $\hat{h}_p^0 = g_{s1} \times Q_{peak} \ln\left(\frac{1}{1-\epsilon}\right)$, respectively. However, in both cases, there is no closed form of the second integral in the capacity expression (16) which may be evaluated numerically. Nevertheless, asymptotic analysis when $P_{avg}$ (respectively $P_{peak}$) is sufficiently large (no budget constraint) or the outage constraint is not effective ($Q_{peak}$ is sufficiently high, for instance), are provided below:

$$\lim_{P_{avg}\to\infty} C = \frac{Q_{eq}\ln(Q_{eq})}{Q_{eq}-1}, \quad (41)$$

where $Q_{eq} = Q_{peak}$ in case of interference constraint and $Q_{eq} = Q_{peak}\ln(1/(1-\epsilon))$ in case of SI outage constraint. Whereas the capacity at infinite outage constraint is given by (38), and is independent of the channel estimation quality.

- Peak transmit-power

In this case, the power profile (19) and the ergodic capacity (20) can be obtained using (21) or (22) for interference outage or SI outage constraints, respectively. At high peak power constraint, it can be shown that

$$\lim_{P_{peak}\to\infty} C = \frac{Q_{eq}\ln(Q_{eq})}{Q_{eq}-1} \quad (42)$$

whereas, the high interference outage constraint ergodic capacity is again independent of the channel estimation quality and is equal to the corresponding one where no ST-PR CSI is available (c.f. Table I).

## VI. Numerical Results

In this section, numerical results are provided for Rayleigh fading channels as derived in Section V. First come our results for an average transmit-power constraint. Figure 3 depicts the optimum power profile in function of the secondary channel $|h_s|^2$ for a given $\hat{h}_p$ value. As shown in Fig. 3, the power profile has typically two or three regions depending on the interference outage constraint value, $P_{\mathbf{X}|\hat{\mathbf{h}}_p}(\epsilon)$. For a relatively high $P_{\mathbf{X}|\hat{\mathbf{h}}_p}(\epsilon)$, the power profile is similar to a water-filling as discussed in [26]; whereas, a low $P_{\mathbf{X}|\hat{\mathbf{h}}_p}(\epsilon)$ value limits the power profile even when the secondary link, $|h_s|^2$ is "good". This limitation behavior affects also the ergodic capacity, no matter how the channel estimation quality is, as shown in Fig. 4. However, at low-power regime ($P_{avg} \to 0$), the ergodic capacity is insensitive to the interference outage constraint and equal capacity is achieved regardless of $P_{\mathbf{X}|\hat{\mathbf{h}}_p}(\epsilon)$ and $\sigma_p^2$. The capacity at infinite interference outage constraint and at infinite average power constraint are also shown in Fig. 4 as performance limits. Interestingly, the first limit is achieved at low-power regime. For instance, when $Q_{peak} = 10$ and $\epsilon = \epsilon_1 = 4.2\%$, so that $\frac{Q_{peak}}{\ln(1/\epsilon_1)} = 5$ dB, then the ergodic capacity with and without interference constraint is the same for $P_{avg}$ below 2 dB, irrespectively to $\sigma_p^2$. The second limit is achieved at relatively high $P_{avg}$ values and increasingly with respect to the channel estimation quality. In Fig. 5, the capacity loss percentage defined as the ratio of the difference between the ergodic capacity under perfect cross link CSI ($\sigma_p^2 \to 0$) and the same capacity where only a noisy cross link CSI with error variance $\sigma_p^2$ is available at the ST, over the capacity under perfect cross link CSI, is plotted for different $P_{avg}$ values. Figure 5 confirms our previous observations: At low $P_{avg} = 0$ dB, the capacity loss is equal to zero and the interference constraint has no effect on the secondary performance; at $P_{avg} = 6$ dB, the cross link CSI uncertainty impacts the secondary capacity and the loss may be over 15% for a poor CSI quality ($\sigma_p^2 \geq 0.85$). However, it is interesting to note that an average CSI quality ($\sigma_p^2 \leq 0.5$) is enough to contain the capacity loss (less than 6%); On the contrary, at $P_{avg} = 10$ dB, the cross link CSI quality is detrimental since the capacity loss may reach up to 40% and one requires a "very good" CSI quality to reduce the capacity loss. For instance, even with $\sigma_p^2 = 0.1$, the capacity loss is more than 10%. This is in fact expected since at high power regime, the capacity is dictated by the interference constraint only, as stipulated by (16).

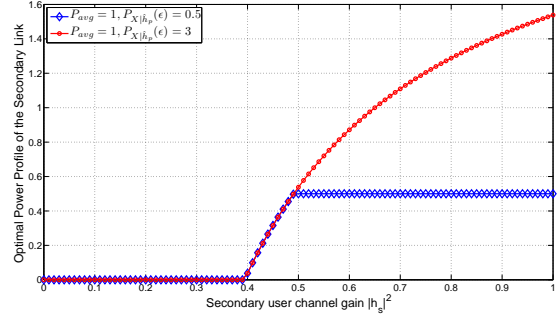

Fig. 3. Optimum power profile of the secondary link for a given $\hat{h}_p$ and under average transmit-power constraint, when $P_{avg} \leq \mathop{\mathrm{E}}_{\hat{\mathbf{h}}_p}\left[P_{|\mathbf{h}_p|^2|\hat{h}_p}(\epsilon)\right]$.

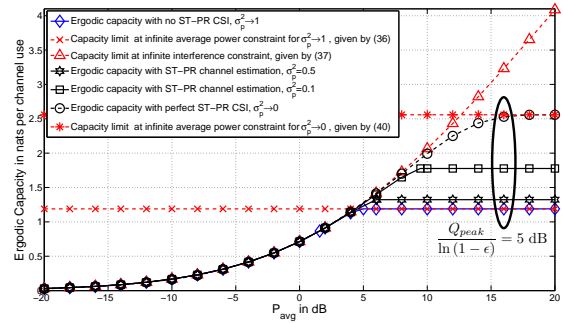

Fig. 4. Ergodic capacity of the secondary link under average transmit-power and interference outage constraints, for different channel estimation error variance values, $\sigma_p^2$.

In order to display results for SI outage constraint, we set $\frac{P_{pp}}{\lambda_{th}} = Q_{peak} = 10$ and $\epsilon = \epsilon_2 = 24\%$ in Fig. 6, so that the interference outage and SI outage constraints are equal in the no ST-PR CSI case. That is, $\frac{1}{\ln(1/\epsilon_1)}Q_{peak} = \frac{\epsilon_2}{1-\epsilon_2}Q_{peak} = 5$ dB, and hence the corresponding ergodic capacities of the secondary user under interference outage and SI outage constraints are equal. Note that to achieve equal ergodic capacity in this case, $\epsilon_2$ must be bigger than $\epsilon_1$ (a higher SI outage should be tolerated), which implies that at no ST-PR CSI case, SI outage constraint is more restrictive than interference outage constraint, from a capacity perspective, when $\frac{P_{pp}}{\lambda_{th}} = Q_{peak}$. Evidence of this can be seen by first noting that the $\epsilon_1 - \epsilon_2$ region $\mathcal{R}_1$ defined by:

$$\mathcal{R}_1 = \left\{(\epsilon_1, \epsilon_2) \mid \ln\frac{1}{\epsilon_1} \geq \frac{1}{\epsilon_2} - 1\right\}, \quad (43)$$

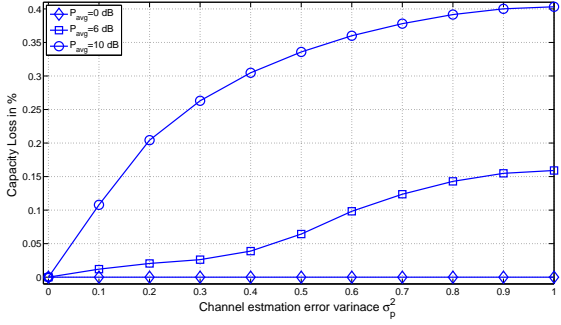

Fig. 5. Capacity Loss percentage of the secondary link under average transmit-power and interference outage constraints, versus channel estimation error variance $\sigma_p^2$, for different $P_{avg}$ values.

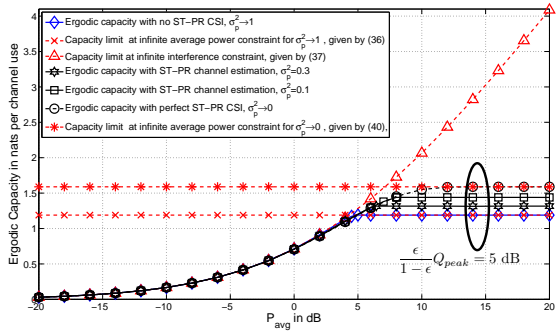

Fig. 6. Ergodic capacity of the secondary link under average transmit-power and SI outage constraints, for different channel estimation error variance values, $\sigma_p^2$.

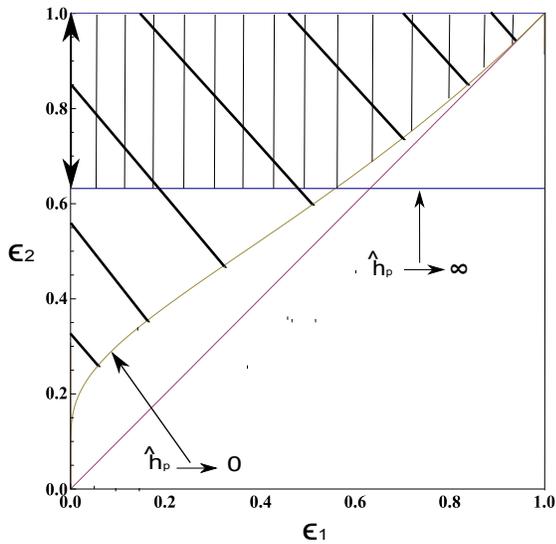

Fig. 7. $\epsilon_1 - \epsilon_2$ regions where the secondary link capacity under SI outage constraint is bigger than the one under interference outage for $\sigma_p^2 = 1$ in oblique lines, for $\sigma_p^2 = 0$ coinciding with y-axis such that $\epsilon_2 \geq 0.63$, and for $0 < \sigma_p^2 < 1$ in vertical lines.

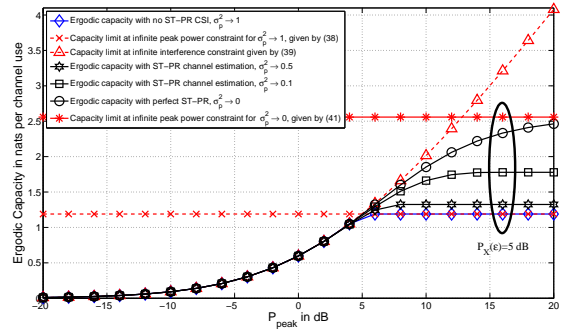

Fig. 8. Ergodic capacity of the secondary link under peak transmit-power constraint and for different channel estimation error variance values, $\sigma_p^2$.

corresponds to the region where the secondary user capacity under SI outage constraint $\epsilon_2$ is bigger than the one under interference outage constraint $\epsilon_1$, in the non ST-PR CSI scenario. Then using the inequality $x \geq \exp\left(1 - \frac{1}{x}\right)$, for all $x > 0$, (43) directly implies that $\epsilon_1 \leq \epsilon_2$. The fact that a higher SI outage is to be tolerated to achieve equal ergodic capacity than the one under interference outage constraint can also be seen for the perfect ST-PR CSI case, where $\epsilon_1 = 0$, by equalizing (21) and (22), which results in $\epsilon_1 - \epsilon_2$ region $\mathcal{R}_2$ defined by:

$$\mathcal{R}_2 = \{(\epsilon_1, \epsilon_2) \mid \epsilon_1 = 0, \epsilon_2 \geq 0.63\}, \quad (44)$$

and thus, all SI outage levels below this value, provide an ergodic capacity smaller than the one corresponding to an interference outage constraint. Only if the primary user is ready to accept an SI outage $\epsilon_2 \geq 63\%$, hence sacrificing its error probability performance, the secondary link would be able to achieve a higher ergodic capacity than the one under interference outage constraint. Furthermore, as shown in Fig. 6, even for such a high SI outage level ($\epsilon_2 = 24\%$), the secondary link ergodic capacity is smaller than the one displayed in Fig. 4 for equal channel estimation quality. When $0 < \sigma_p^2 < 1$, this fact can similarly be explained by noting that the secondary user capacity under SI outage is bigger than the one under interference outage iff $(\epsilon_1, \epsilon_2) \in \mathcal{R}_3$, where $\mathcal{R}_3$ is defined by:

$$\mathcal{R}_3 = \left\{(\epsilon_1, \epsilon_2) \mid P_{|\mathbf{h}_p|^2|\hat{\mathbf{h}}_p}(\epsilon_1) \leq P_{\beta|\hat{\mathbf{h}}_p}(\epsilon_2)\right\}, \quad (45)$$

for all $\hat{h}_p$. By letting $\hat{h}_p$ tends toward 0, it can be verified that $F^{-1}_{|\mathbf{h}_p|^2|\hat{\mathbf{h}}_p}(x) \to \sigma_p^2 \ln \frac{1}{x}$ and that $F^{-1}_{\beta|\hat{\mathbf{h}}_p}(x) \to \sigma_p^2 \ln \frac{x}{1-x}$. Hence, as $\hat{h}_p$ tends toward 0, (45) implies (43), which itself implies that $\epsilon_1 \leq \epsilon_2$ as proved above. Figure 7 illustrates the $\epsilon_1 - \epsilon_2$ regions $\mathcal{R}_1$, $\mathcal{R}_2$ and $\mathcal{R}_3$ defined by (43), (44) and (45), respectively.

Then come our results for a peak transmit-power constraint. Although the interference outage constraint limits the capacity at high-power regime, it has again no effect at low-power regime and no-interference performance is achieved as shown in Fig. 8. Finally, it is to be mentioned that at low power regime, the peak constraint is more stringent than the average constraint from a capacity point of view, in agreement with [33]. For instance, at a given interference outage constraint, say $P_X(\epsilon) = 5$ dB, the ergodic capacity provided by a peak transmit-power constraint $P_{peak} = 0$ dB is equal to 0.6 npcu, whereas for an average transmit-power $P_{avg} = 0$ dB, the ergodic capacity is equal to 0.71 npcu. This gap diminishes remarkably at high power regime where the outage constraint dictates the power profile.

## VII. Conclusion

A spectrum-sharing communication with ST-PR channel estimation at the secondary user has been addressed. The optimum power profile and the ergodic capacity have been derived for a class of fading channels with respect to an average or a peak transmit-power, along with more realistic interference outage constraints. The impact of channel estimation quality on the ergodic capacity has been highlighted. In all cases, asymptotic analysis has been discussed in order to provide a better understanding of the performance limits of a spectrum-sharing protocol. Our framework generalizes and encompasses several existent results.

## Appendix A
### Derivation of the optimum power profile given by (9) and equivalently by (11), (12) and (13)

To solve the optimization problem formulated by (5), (6) and (8), we first consider the set, say $\mathcal{A}$, of all $h_s$ and $\hat{h}_p$ values such that (8) is satisfied. That is, $\mathcal{A} = \left\{h_s, \hat{h}_p : P(h_s, \hat{h}_p) \leq P_{|\mathbf{h}_p|^2|\hat{h}_p}(\epsilon)\right\}$. Our optimization can thus be formulated by (5), (6), over the set $\mathcal{A}$. Since (5) is concave, (6) is linear and $\mathcal{A}$ is convex, then the new equivalent optimization problem is concave over $\mathcal{A}$. The solution of (5) and (6) over all channel realizations $h_s$ and $\hat{h}_p$ is known to be a waterfilling given by $P(h_s, \hat{h}_p) = \left[\frac{1}{\lambda} - \frac{1}{|h_s|^2}\right]^+$. However, since we are now optimizing only over $\mathcal{A}$, the optimal power is capped by $P_{|\mathbf{h}_p|^2|\hat{h}_p}(\epsilon)$, which leads to (9).

Departing from (9), we show that the optimum power profile is given by (11), (12) and (13). First, note that since the optimization problem given by (5), (6) and (7) is convex with a feasible point, then $\lambda$ in (9) can be found by solving the (concave) Lagrange dual problem defined by [34]:

$$\min_{\lambda \geq 0} g(\lambda), \quad (46)$$

where $g(\lambda)$ is given by:

$$g(\lambda) = \mathop{\mathbb{E}}_{\mathbf{h}_s, \hat{\mathbf{h}}_p}\left[\ln\left(1 + P^*(\mathbf{h}_s, \hat{\mathbf{h}}_p) \cdot |h_s|^2\right)\right] - \lambda\left(\mathop{\mathbb{E}}_{\mathbf{h}_s, \hat{\mathbf{h}}_p}\left[P^*(\mathbf{h}_s, \hat{\mathbf{h}}_p)\right] - P_{avg}\right)$$

and where $P^*(\mathbf{h}_s, \hat{\mathbf{h}}_p)$ is the optimal solution given by (9). Since $g(\lambda)$ is concave with a feasible minimum, then, we have:

- **C1**: If $\frac{\partial g}{\partial \lambda}|_{\lambda=0} \geq 0$, then $\min_{\lambda \geq 0} g(\lambda) = g(0)$,
- **C2**: otherwise, $\min_{\lambda \geq 0} g(\lambda) = g(g_{s1})$, for some $g_{s1} > 0$.

Now, it can be easily verified that

$$\frac{\partial g}{\partial \lambda} = \left(P_{avg} - \mathop{\mathbb{E}}_{\mathbf{h}_s, \hat{\mathbf{h}}_p}\left[P^*(\mathbf{h}_s, \hat{\mathbf{h}}_p)\right]\right),$$

and thus

$$\frac{\partial g}{\partial \lambda}|_{\lambda=0} = \left(P_{avg} - \mathop{\mathbb{E}}_{\hat{\mathbf{h}}_p}\left[P_{|\mathbf{h}_p|^2|\hat{h}_p}(\epsilon)\right]\right). \quad (47)$$

Combining (47), condition **C1** and condition **C2**, together with (9) yield (11), (12) and (13). It remains to prove (14). If **C2** holds, then by the complimentary slackness condition [34], the average power constraint is satisfied with equality, and we have:

$$P_{avg} = \mathop{\mathbb{E}}_{\mathbf{h}_s, \hat{\mathbf{h}}_p}\left[P^*(\mathbf{h}_s, \hat{\mathbf{h}}_p)\right]$$

$$= \mathop{\mathbb{E}}_{\hat{\mathbf{h}}_p}\left[\mathop{\mathbb{E}}_{\mathbf{h}_s|\hat{\mathbf{h}}_p}\left[P^*(\mathbf{h}_s, \hat{\mathbf{h}}_p) \mid \hat{\mathbf{h}}_p\right]\right]$$

$$= \mathop{\mathbb{E}}_{\hat{\mathbf{h}}_p \in \mathcal{S}_{g_{s1}}}\left[\mathop{\mathbb{E}}_{\mathbf{h}_s|\hat{\mathbf{h}}_p}\left[P^*(\mathbf{h}_s, \hat{\mathbf{h}}_p) \mid \hat{\mathbf{h}}_p \in \mathcal{S}_{g_{s1}}\right]\right]$$

$$+ \mathop{\mathbb{E}}_{\hat{\mathbf{h}}_p \in \bar{\mathcal{S}}_{g_{s1}}}\left[\mathop{\mathbb{E}}_{\mathbf{h}_s|\hat{\mathbf{h}}_p}\left[P^*(\mathbf{h}_s, \hat{\mathbf{h}}_p) \mid \hat{\mathbf{h}}_p \in \bar{\mathcal{S}}_{g_{s1}}\right]\right] \quad (48)$$

Now, the outer expectation in the second term on the right hand side of (48) is over all $\hat{\mathbf{h}}_p \in \bar{\mathcal{S}}_{g_{s1}}$ and hence the power profile is given by (13). Therefore, we can compute the second term in (48) as follows:

$$\mathop{\mathbb{E}}_{\hat{\mathbf{h}}_p \in \bar{\mathcal{S}}_{g_{s1}}}\left[\mathop{\mathbb{E}}_{\mathbf{h}_s|\hat{\mathbf{h}}_p}\left[P^*(\mathbf{h}_s, \hat{\mathbf{h}}_p) \mid \hat{\mathbf{h}}_p \in \bar{\mathcal{S}}_{g_{s1}}\right]\right] = \mathop{\mathbb{E}}_{\hat{\mathbf{h}}_p \in \bar{\mathcal{S}}_{g_{s1}}}\left[\mathop{\mathbb{E}}_{\mathbf{h}_s|\hat{\mathbf{h}}_p}\left[\frac{1}{g_{s1}} - \frac{1}{|h_s|^2}\right]^+\right]$$

$$= \mathop{\mathbb{E}}_{\hat{\mathbf{h}}_p \in \bar{\mathcal{S}}_{g_{s1}}}\left[\mathop{\mathbb{E}}_{\mathbf{h}_s}\left[\frac{1}{g_{s1}} - \frac{1}{|h_s|^2}\right]^+\right] \quad (49)$$

$$= \mathop{\mathbb{E}}_{\hat{\mathbf{h}}_p \in \bar{\mathcal{S}}_{g_{s1}}}\left[G(g_{s1})\right], \quad (50)$$

where (49) follows from the independence of $\mathbf{h}_s$ and $\hat{\mathbf{h}}_p$. On the contrary, the outer expectation in the first term on the right hand side of (48) is over all $\hat{\mathbf{h}}_p \in \mathcal{S}_{g_{s1}}$ and hence the power profile is given by (12). Therefore, we can compute the first term in (48) as shown at the bottom of the page. Now,

$$\mathop{\mathbb{E}}_{\hat{\mathbf{h}}_p \in \mathcal{S}_{g_{s1}}}\left[\mathop{\mathbb{E}}_{\mathbf{h}_s|\hat{\mathbf{h}}_p}\left[P^*(\mathbf{h}_s, \hat{\mathbf{h}}_p) \mid \hat{\mathbf{h}}_p \in \mathcal{S}_{g_{s1}}\right]\right] = \mathop{\mathbb{E}}_{\hat{\mathbf{h}}_p \in \mathcal{S}_{g_{s1}}}\left[\mathop{\mathbb{E}}_{|h_s|^2 \in [g_{s1}, g_{s2})}\left[\frac{1}{g_{s1}} - \frac{1}{|h_s|^2}\right]^+ + \mathop{\mathbb{E}}_{|h_s|^2 \geq g_{s2}}\left[P_{|\mathbf{h}_p|^2|\hat{h}_p}(\epsilon)\right]\right]$$

$$= \mathop{\mathbb{E}}_{\hat{\mathbf{h}}_p \in \mathcal{S}_{g_{s1}}}\left[G(g_{s1}) - \mathop{\mathbb{E}}_{|h_s|^2 > g_{s2}}\left[\frac{1}{g_{s1}} - \frac{1}{|h_s|^2} - P_{|\mathbf{h}_p|^2|\hat{h}_p}(\epsilon)\right]^+\right]$$

$$= \mathop{\mathbb{E}}_{\hat{\mathbf{h}}_p \in \mathcal{S}_{g_{s1}}}\left[G(g_{s1}) - G(g_{s2})\right],$$

$$= \mathop{\mathbb{E}}_{\hat{\mathbf{h}}_p \in \mathcal{S}_{g_{s1}}}\left[G(g_{s1})\right] - \mathop{\mathbb{E}}_{\hat{\mathbf{h}}_p \in \mathcal{S}_{g_{s1}}}\left[G(g_{s2})\right] \quad (51)$$

gathering (48), (50) and (51), we obtain:

$$P_{avg} = \mathop{\mathbb{E}}_{\hat{\mathbf{h}}_p}[G(g_{s1})] - \mathop{\mathbb{E}}_{\hat{\mathbf{h}}_p \in \mathcal{S}_{g_{s1}}}[G(g_{s2})]$$
$$= G(g_{s1}) - \mathop{\mathbb{E}}_{\hat{\mathbf{h}}_p \in \mathcal{S}_{g_{s1}}}[G(g_{s2})], \quad (52)$$

where (52) follows because $g_{s1}$ depends only on $P_{avg}$. Using the fact that the function $K(x)$ in (15) is invertible on $(0, P_{|\mathbf{h}_p|^2|\hat{h}_p}(\epsilon)]$, (14) follows immediately. Finally, since $K(x)$ is monotonically decreasing, then $g_{s1} < \frac{1}{P_{|\mathbf{h}_p|^2|\hat{h}_p}(\epsilon)}$ is equivalent to $P_{avg} > G\left(1/P_{|\mathbf{h}_p|^2|\hat{h}_p}(\epsilon)\right)$.

ACKNOWLEDGMENT

The authors would like to thank Jose Roberto Ayala and Lokman Sboui for their constructive comments.

**Zouheir Rezki** (S'01, M'08) was born in Casablanca, Morocco. He received the Diplome d'Ingénieur degree from the École Nationale de l'Industrie Minérale (ENIM), Rabat, Morocco, in 1994, the M.Eng. degree from École de Technologie Supérieure, Montreal, Québec, Canada, in 2003, and the Ph.D. degree from École Polytechnique, Montreal, Québec, in 2008, all in electrical engineering. From October 2008 to September 2009, he was a postdoctoral research fellow with Data Communications Group, Department of Electrical and Computer Engineering, University of British Columbia. He is now a postdoctoral research fellow at King Abdullah University of Science and Technology (KAUST), Thuwal, Mekkah Province, Saudi Arabia. His research interests include: performance limits of communication systems, cognitive and sensor networks, physical-layer security, and low-complexity detection algorithms.

**Mohamed-Slim Alouini** (S'94, M'98, SM'03, F'09) was born in Tunis, Tunisia. He received the Ph.D. degree in electrical engineering from the California Institute of Technology (Caltech), Pasadena, CA, USA, in 1998. He was with the department of Electrical and Computer Engineering of the University of Minnesota, Minneapolis, MN, USA, then with the Electrical and Computer Engineering Program at the Texas A&M University at Qatar, Education City, Doha, Qatar. Since June 2009, he has been a Professor of Electrical Engineering in the Division of Physical Sciences and Engineering at KAUST, Saudi Arabia., where his current research interests include the design and performance analysis of wireless communication systems.